\documentclass[%
 twocolumn,
 amsmath,amssymb,
 aps,
 prb,
 10pt,
 floatfix,
]{revtex4-2}

\usepackage{graphicx}
\usepackage{dcolumn}
\usepackage{bm}
\usepackage{hyperref}

\begin{document}

\title{Efficient Interpolation of the Bethe--Salpeter Hamiltonian at Arbitrary Momenta}

\author{Stefano Mocatti}
 \email{stefano.mocatti@unitn.it}
 \affiliation{Department of Physics, University of Trento, Via Sommarive 14, 38123 Povo, Italy}%

\author{Xiangzhou Zhu}
 \affiliation{Department of Physics, University of Trento, Via Sommarive 14, 38123 Povo, Italy}%

\author{Giulio Volpato}
 \affiliation{Department of Physics, University of Trento, Via Sommarive 14, 38123 Povo, Italy}%

 \author{Pierluigi Cudazzo}
 \affiliation{Department of Physics, University of Trento, Via Sommarive 14, 38123 Povo, Italy}%

\author{Matteo Calandra}
 \affiliation{Department of Physics, University of Trento, Via Sommarive 14, 38123 Povo, Italy}%

\author{Giovanni Marini}
 \affiliation{Department of Physics, University of Trento, Via Sommarive 14, 38123 Povo, Italy}%

\date{\today}

\begin{abstract}
We develop an efficient reciprocal-space method to reconstruct the static Bethe--Salpeter (BSE) Hamiltonian at arbitrary crystal momenta from coarse first-principles grids. Bloch states are interpolated within an optimally smooth Wannier subspace, while the screened Coulomb interaction is reconstructed from the dielectric response. This decouples the electron--hole transition momenta and transferred momentum from the underlying \textit{ab initio} sampling. We benchmark the method on monolayer hBN, monolayer MoS$_2$, and bulk GaAs against explicit \textit{ab initio} calculations, finding excellent agreement for dense-grid Bloch states, dielectric screening, and BSE optical spectra. Furthermore, our cost analysis estimates speedups of up to $22\times$ on the grids considered here. Arbitrary transferred-momentum sampling gives direct access to the long-wavelength exciton dispersion: in hBN and MoS$_2$, we resolve the nonanalytic behaviour induced by long-range electron--hole exchange at momenta well below the spacing of practical uniform grids. For bulk GaAs, locally refined transition-momentum sampling yields a converged binding energy for the threefold-degenerate lowest exciton, in close agreement with experiment and reference calculations, using far fewer BSE points than conventional uniform meshes. As the present framework provides exciton energies and amplitudes at arbitrary crystal momenta, it opens the way to incorporating \textit{ab initio} excitonic effects into nonlocal electron--phonon vertices and to extensions towards exciton--phonon interactions and nonequilibrium exciton dynamics.
\end{abstract}

\maketitle

\section{Introduction}

The Bethe--Salpeter equation (BSE) provides a standard first-principles description of neutral excitations in solids, including optical spectra, exciton binding energies, and correlated electron--hole states~\cite{Strinati_1988,Albrecht_1998,Rohlfing_2000,Onida_2002}. Its predictive accuracy, however, comes with a substantial computational cost. The electron--hole Hamiltonian couples transitions throughout the Brillouin zone, and converged calculations often require dense momentum sampling, particularly for spatially extended excitons~\cite{Fuchs_2008,Alvertis_2023}. The screened Coulomb interaction introduces a further sampling bottleneck, since its momentum dependence can vary rapidly in the long-wavelength limit, especially in low-dimensional systems~\cite{Huser_2013,Qiu_2016}.

The need for flexible momentum sampling becomes even more apparent away from the optical limit. Standard first-principles implementations rely on regular electronic momentum meshes, so a finite-momentum BSE calculation can be performed directly only when the electronic states required for a given transferred momentum $\mathbf Q$, corresponding to the total crystal momentum of the electron--hole pair, are represented on those grids. An arbitrary $\mathbf Q$ will generally not satisfy this condition, and accessing it may require a new, typically denser, underlying calculation. This becomes particularly restrictive close to $\Gamma$ in two-dimensional systems, where long-range electron--hole exchange gives rise to a nonanalytic exciton dispersion and distinct longitudinal and transverse branches~\cite{Andreani_1990,Qiu_2015,Cudazzo_2016}. Resolving this regime requires transferred momenta well below the spacing of typical \textit{ab initio} meshes, as recently demonstrated by the direct observation of a linear exciton dispersion in monolayer hBN~\cite{Liu_2026,Hong_2026}.

Dense interpolation in reciprocal space is already widely used in first-principles calculations. Maximally localised Wannier functions~\cite{Marzari_1997,Souza_2001,Marzari_2012} provide a smooth representation of Bloch states from which band structures can be reconstructed on dense momentum grids. The same framework was subsequently extended to electron--phonon matrix elements~\cite{Giustino_2007,Calandra_2010}, while more recent work has shown that reciprocal-space quantities can be interpolated directly within an optimally smooth Wannier subspace, without an intermediate transformation to real space~\cite{Volpato_2025}. These developments now underpin dense-grid first-principles calculations of electron--phonon coupling, transport, spectroscopy, and carrier dynamics~\cite{Lee_2023,Zhou_2021,Cepellotti_2022,Marini_2024}.

A common smooth electronic representation becomes especially important when several interactions are treated within the same computational framework. Transport and nonequilibrium calculations require electronic states and interaction vertices to be evaluated on compatible dense momentum grids. This is the case, for example, in coupled Boltzmann treatments including both electron--phonon and electron--electron scattering~\cite{Macheda_2026}, and in real-time descriptions of carrier dynamics in which both scattering channels are treated simultaneously~\cite{Marini_2013,Sangalli_2015,Perfetto_2023,Stefanucci_2024,Mocatti_2026}. Beyond interpolation, a common Wannier representation allows matrix elements associated with different interactions to be combined consistently.

The same requirement arises when excitons are coupled to lattice degrees of freedom. First-principles exciton--phonon frameworks combine BSE amplitudes at different transferred momenta with electron--phonon matrix elements to describe scattering and relaxation~\cite{Chen_2020,Antonius_2022}, phonon-assisted absorption and emission~\cite{Paleari_2019}, dynamical spectral renormalisation and phonon sidebands~\cite{Cudazzo_2020}, exciton lifetimes and optical linewidths~\cite{Chan_2023}, and phonon-driven exciton dynamics and thermalisation in real time~\cite{Stefanucci_2025,Betancur_2026}. The symmetry properties of finite-momentum excitons further constrain these processes through selection rules for exciton--phonon coupling~\cite{Nalabothula_2026}. Recent formulations have also addressed the consistent combination of excitonic and electron--phonon quantities within a common electronic representation~\cite{Alvertis_2024,Marini_Calandra_Cudazzo_2024}. More broadly, the interplay between electronic correlations and lattice dynamics motivates a common treatment of the corresponding interaction vertices~\cite{Guandalini_2025}. The ability to reconstruct the BSE interaction at arbitrary momenta within the same smooth electronic representation as other interaction vertices is therefore relevant not only for equilibrium excitonic properties, but also for exciton transport and nonequilibrium dynamics.

Making such calculations practical also requires reducing the cost of constructing and solving the underlying many-body problem. Compact representations of the polarisation based on localised orbitals and formulations avoiding explicit sums over empty states have been developed for $GW$ calculations~\cite{Umari_2009,Umari_2010}. Maximally localised Wannier functions have been used in large-scale $GW$--BSE calculations to reduce their formal scaling~\cite{Marsili_2017} and to reduce the cost of real-time HSEX calculations of excitonic and nonlinear optical response~\cite{Cistaro_2026}. Locality has further been exploited directly at the excitonic level, yielding sparse real-space BSE Hamiltonians and efficient evaluations of the electron--hole interaction~\cite{Merkel_2024}. Lanczos-chain and response-based approaches avoid explicit diagonalisation of the full response problem~\cite{Rocca_2008,Baroni_2010}, while density-matrix~\cite{Rocca_2010,Rocca_2012} and finite-field formulations~\cite{Nguyen_2019} reduce or eliminate explicit summations over virtual states and dielectric matrices.

Dense momentum sampling nevertheless remains a distinct computational bottleneck in BSE calculations, and several strategies have been developed to alleviate it. Early reciprocal-space schemes interpolated the electron--hole kernel from a coarse to a finer momentum mesh~\cite{Rohlfing_1998}, with later formulations evaluating the interpolated matrix elements on the fly to reduce memory and computational cost in iterative BSE solvers~\cite{Gillet_2016}. Double-grid approaches instead retain the expensive kernel on the coarse mesh and use a finer auxiliary grid for the single-particle contribution and optical response~\cite{Kammerlander_2012,Alliati_2022}. Nonuniform samplings, including locally refined and hybrid meshes, concentrate momentum points in the regions most relevant to the exciton wave function~\cite{Fuchs_2008,Alvertis_2023}, while interpolation of the screened interaction has been used to treat its sharp momentum dependence near the long-wavelength limit in reduced-dimensional systems~\cite{Guandalini_2023}. At a different level, maximally localised exciton Wannier functions~\cite{Haber_2023}, constructed from Bloch excitons, allow exciton energies and wave functions to be interpolated as functions of the transferred momentum $\mathbf Q$.

In this work, we develop a reciprocal-space framework in which both the Bloch states and screened interaction can be reconstructed independently at arbitrary momenta. The approach is implemented in the code \textsc{EPIq}~\cite{Marini_2024}. Bloch states are interpolated within an optimally smooth Wannier subspace~\cite{Volpato_2025}, while the screened interaction is reconstructed from an interpolated dielectric response~\cite{Guandalini_2023}. This decouples the BSE sampling from the underlying first-principles grids, allowing both the transition-momentum grid $\mathbf k$ and the transferred momentum $\mathbf Q$ to be chosen independently without repeating the expensive \textit{ab initio} calculation. At the same time, all electronic quantities remain tied to a common smooth representation, allowing the resulting exciton wave functions to be combined consistently with other interaction vertices defined in the same electronic representation.

We benchmark the method for monolayer hBN and MoS$_2$ and bulk GaAs, first at the level of the interpolated Bloch states and dielectric response and then through BSE optical spectra on dense target meshes. We also quantify the computational gain obtained by replacing dense first-principles sampling with interpolation. Finally, we demonstrate the flexibility of the target BSE sampling with two applications: resolving the long-wavelength exciton dispersions of hBN and MoS$_2$ at arbitrary transferred momenta, and converging the binding energies of weakly bound GaAs excitons using locally refined transition-momentum meshes.

\section{Theoretical framework}

We reconstruct the static BSE Hamiltonian at arbitrary exciton momenta by interpolating two quantities: the reciprocal-space components of the Bloch states in an optimally smooth Wannier representation \cite{Volpato_2025} and the static dielectric screening.  The direct and exchange electron--hole kernels are then calculated and the BSE Hamiltonian solved.

Rydberg atomic units (a.u.) are used throughout unless otherwise stated,
\begin{equation}
    \hbar = 2m_e = \frac{e^2}{2} = 4\pi\epsilon_0 = 1,
\end{equation}
with $m_e$ the electron mass, $e$ the magnitude of the elementary charge, and $\epsilon_0$ the vacuum permittivity.

\subsection{Finite-momentum Bethe--Salpeter Hamiltonian}
\label{subsec:bse_formalism}

The Bloch functions are written as
\begin{equation}
\label{eq:bloch_wavefunction}
\psi_{\mathbf kn}(\mathbf r,\sigma)
=
\frac{1}{\sqrt{N}}
e^{i\mathbf k\cdot\mathbf r}
u_{\mathbf kn}(\mathbf r,\sigma),
\end{equation}
where $\sigma$ labels the spinor component and $N$ is the number of primitive cells in the Born--von K\'arm\'an (BvK) supercell. The Bloch states $\psi_{\mathbf kn}$ are normalised over the BvK volume $\mathcal V=N\Omega$, with $\Omega$ the primitive-cell volume, while the periodic parts $u_{\mathbf kn}$ are normalised over $\Omega$. For a uniform BZ sampling, $N$ coincides with the number $N_k$ of crystal momenta in the full BZ.

A neutral electron--hole excitation with transferred momentum $\mathbf Q$, corresponding to the total crystal momentum of the electron--hole pair, is represented in the transition basis
\begin{equation}
\label{eq:transition_basis}
|mn\mathbf k;\mathbf Q\rangle
=
\hat{c}^{\dagger}_{\mathbf k+\mathbf Qn}
\hat{c}_{\mathbf km}|0\rangle ,
\end{equation}
where $\hat{c}_{\mathbf kn}$ ($\hat{c}^{\dagger}_{\mathbf kn}$) annihilates (creates) an electron in Bloch state $n$ at momentum $\mathbf k$, with single-particle energy $\varepsilon_{\mathbf kn}$. The momentum $\mathbf k$ therefore labels the electron--hole transition, with the hole at $\mathbf k$ and the electron at $\mathbf k+\mathbf Q$.

We consider insulating and semiconducting systems within the resonant Tamm--Dancoff approximation. The transition space is restricted to excitations from occupied valence states $v$ to unoccupied conduction states $c$, corresponding to $m=v$ and $n=c$ in Eq.~\eqref{eq:transition_basis}. Neglecting the coupling between resonant and anti-resonant sectors then yields a Hermitian eigenvalue problem~\cite{Onida_2002,Sander_2015}. For a given transferred momentum $\mathbf Q$, the BSE Hamiltonian is
\begin{equation}
\label{eq:bse_hamiltonian}
H^{\mathrm{BSE}}(\mathbf Q)
=
H^0(\mathbf Q)
+
K^d(\mathbf Q)
+
K^x(\mathbf Q),
\end{equation}
with independent-particle contribution
\begin{equation}
\label{eq:bse_independent}
H^0_{vc\mathbf k,v'c'\mathbf k'}(\mathbf Q)
=
\left[
\varepsilon_{\mathbf k+\mathbf Qc}
-
\varepsilon_{\mathbf kv}
\right]
\delta_{vv'}
\delta_{cc'}
\delta_{\mathbf k\mathbf k'} ,
\end{equation}
and screened direct and bare exchange kernels $K^d$ and $K^x$, respectively~\cite{Rohlfing_2000,Onida_2002}. The momenta $\mathbf k$ and $\mathbf k'$ label the two electron--hole transitions coupled by a matrix element of the BSE Hamiltonian.

The interaction kernels require transition densities between Bloch states,
\begin{equation}
\label{eq:rho_bloch}
\rho^{\mathbf G}_{mn}(\mathbf k,\mathbf q)
=
\sum_{\sigma}
\int_{\mathcal V}d\mathbf r\,
\psi^{*}_{\mathbf k+\mathbf qm}(\mathbf r,\sigma)
e^{i(\mathbf q+\mathbf G)\cdot\mathbf r}
\psi_{\mathbf kn}(\mathbf r,\sigma),
\end{equation}
or, equivalently, in terms of the cell-periodic parts,
\begin{equation}
\label{eq:rho_periodic}
\rho^{\mathbf G}_{mn}(\mathbf k,\mathbf q)
=
\sum_{\sigma}
\int_{\Omega}d\mathbf r\,
u^{*}_{\mathbf k+\mathbf qm}(\mathbf r,\sigma)
e^{i\mathbf G\cdot\mathbf r}
u_{\mathbf kn}(\mathbf r,\sigma).
\end{equation}
Here, $\mathbf k$ and $\mathbf q$ lie in the first Brillouin zone, while $\mathbf G$ is a reciprocal-lattice vector.

The statically screened direct kernel is
\begin{equation}
\label{eq:direct_kernel}
\begin{split}
K^d_{vc\mathbf k,v'c'\mathbf k'}(\mathbf Q)
={}&
-\frac{\sqrt{w_{\mathbf k}w_{\mathbf k'}}}{\Omega}
\sum_{\mathbf G\mathbf G'}
\rho^{\mathbf G}_{cc'}
(\mathbf k'+\mathbf Q,\mathbf q)
\\
&\times
W_{\mathbf G\mathbf G'}(\mathbf q)
\left[
\rho^{\mathbf G'}_{vv'}
(\mathbf k',\mathbf q)
\right]^{*},
\end{split}
\end{equation}
where $w_{\mathbf k}$ and $w_{\mathbf k'}$ are the Brillouin-zone integration weights associated with the transition momenta $\mathbf k$ and $\mathbf k'$. For uniform sampling over the full Brillouin zone, $w_{\mathbf k}=w_{\mathbf k'}=1/N_k$. Here, $\mathbf q=\mathbf k-\mathbf k'$ is the momentum exchanged between the two electron--hole transitions coupled by the direct kernel. It is distinct from the transferred momentum $\mathbf Q$, which is the total crystal momentum of the electron--hole pair. The quantity $W_{\mathbf G\mathbf G'}(\mathbf q)$ is the static screened interaction and is obtained from the inverse dielectric matrix $\epsilon^{-1}_{\mathbf G\mathbf G'}(\mathbf q)$. The dielectric response is evaluated from the independent-particle response within the random-phase approximation, as described in Appendix~\ref{subsec:static_dielectric_screening}.

The bare exchange kernel reads
\begin{equation}
\label{eq:exchange_kernel}
\begin{split}
K^x_{vc\mathbf k,v'c'\mathbf k'}(\mathbf Q)
={}&
\frac{g_s\sqrt{w_{\mathbf k}w_{\mathbf k'}}}{\Omega}
\sum_{\mathbf G}
\rho^{\mathbf G}_{cv}
(\mathbf k,\mathbf Q)
\\
&\times
v_{\mathbf G}(\mathbf Q)
\left[
\rho^{\mathbf G}_{c'v'}
(\mathbf k',\mathbf Q)
\right]^{*},
\end{split}
\end{equation}
where $v_{\mathbf G}(\mathbf Q)$ is the Fourier transform of the bare Coulomb interaction, whose explicit form for different dimensionalities is given in Appendix~\ref{sec:bare_coulomb}. Depending on the response function of interest, the full interaction $v$ in Eq.~\eqref{eq:exchange_kernel} can be replaced by the modified interaction $\bar v$, defined by setting its macroscopic component to zero, $\bar v_{\mathbf G=\mathbf 0}(\mathbf Q)=0$~\cite{Onida_2002}. The full interaction gives the polarisation $\chi$, whereas $\bar v$ is used to obtain the modified polarisation $\bar\chi$ entering the macroscopic optical dielectric function. In the finite-$\mathbf Q$ exciton calculations considered below, we retain the full interaction; in two-dimensional systems, its $\mathbf G=\mathbf 0$ component gives rise to the nonanalytic long-wavelength exciton dispersion~\cite{Qiu_2015,Cudazzo_2016}. In spin-degenerate scalar or collinear singlet calculations, $g_s=2$, while $g_s=1$ when spin is treated explicitly in a non-collinear spinorial formulation. In the conventional spin-conserving BSE, the exchange term vanishes in the triplet channel~\cite{Rohlfing_2000}.

In the following, we show how the BSE Hamiltonian can be constructed on an arbitrary transition-momentum grid and for an arbitrary transferred momentum $\mathbf Q$. The procedure is:
\begin{enumerate}
    \item For each pair of transition momenta $\mathbf k$ and $\mathbf k'$, we interpolate the reciprocal-space components of the Bloch states at the required momenta $\mathbf k$, $\mathbf k+\mathbf Q$, $\mathbf k'$, and $\mathbf k'+\mathbf Q$ within the optimally smooth Wannier subspace~\cite{Volpato_2025}, and construct the corresponding transition densities.
    \item For the resulting exchanged momentum $\mathbf q=\mathbf k-\mathbf k'$, we interpolate the screening response and reconstruct $\epsilon^{-1}_{\mathbf G\mathbf G'}(\mathbf q)$ and $W_{\mathbf G\mathbf G'}(\mathbf q)$. The bare Coulomb interaction is never interpolated and is evaluated directly at the required momentum.
    \item We construct the direct and exchange kernels and assemble the BSE Hamiltonian of Eq.~\eqref{eq:bse_hamiltonian}.
    \item We solve the BSE Hamiltonian to obtain the desired excitonic properties at the chosen transferred momentum $\mathbf Q$.
\end{enumerate}
We now describe these steps in detail.

\section{Interpolation of the BSE Hamiltonian at arbitrary momenta}

\subsection{Wannier interpolation of transition densities}
\label{subsec:wannier_transition_density}

The direct and exchange kernels in Eqs.~\eqref{eq:direct_kernel} and~\eqref{eq:exchange_kernel} require transition densities involving Bloch states at momenta that do not necessarily belong to the original first-principles mesh. We obtain these states by directly interpolating the reciprocal-space components of the Bloch wave functions within the optimally smooth subspace (OSS)~\cite{Volpato_2025}.

The periodic part of a Bloch function is expanded in plane waves as
\begin{equation}
u_{\mathbf kn}(\mathbf r,\sigma)
=
\frac{1}{\sqrt{\Omega}}
\sum_{\mathbf G}
u_{\mathbf kn}(\mathbf G,\sigma)
e^{i\mathbf G\cdot\mathbf r},
\end{equation}
where $u_{\mathbf kn}(\mathbf r,\sigma)$ and $u_{\mathbf kn}(\mathbf G,\sigma)$ denote its real- and reciprocal-space representations, respectively.

Direct interpolation of the coefficients $u_{\mathbf kn}(\mathbf G,\sigma)$ in the band-eigenstate representation is hindered by the absence of a smooth representation of the Bloch eigenstates obtained independently at each crystal momentum. Their phases are arbitrary and, within degenerate or entangled manifolds, the states may also undergo unitary rotations between neighbouring momenta. The smoothness of the Bloch functions in reciprocal space is directly related to the localisation of the corresponding Wannier functions in real space~\cite{Kohn_1959,desCloizeaux_1964,Brouder_2007}.

We therefore use maximally localised Wannier functions (MLWFs)~\cite{Marzari_1997,Souza_2001,Marzari_2012} to construct a smooth electronic representation for interpolation. In the general case of an entangled band manifold, the transformation to the OSS reads
\begin{equation}
u^{\mathrm{OSS}}_{\mathbf kn}(\mathbf G,\sigma)
=
\sum_m
u_{\mathbf km}(\mathbf G,\sigma)
\left[
U^{\mathrm{dis}}(\mathbf k)
U(\mathbf k)
\right]_{mn},
\end{equation}
where $U^{\mathrm{dis}}(\mathbf k)$ defines the disentangled subspace and $U(\mathbf k)$ is the unitary transformation defining the optimally smooth Wannier representation. For an isolated group of bands, $U^{\mathrm{dis}}(\mathbf k)$ is not required. The resulting OSS plane-wave components vary smoothly with $\mathbf k$ and can therefore be interpolated directly in reciprocal space~\cite{Volpato_2025}.

The transition densities must ultimately be constructed in the band-eigenstate representation. At each target momentum, the interpolated OSS states are therefore rotated back to the Hamiltonian eigenstates. The Wannier Hamiltonian is first reconstructed at $\mathbf k$ using either conventional real-space Fourier interpolation~\cite{Marzari_2012} or direct Wannier interpolation in the OSS~\cite{Volpato_2025}, and then diagonalised,
\begin{equation}
\sum_m
H_{nm}(\mathbf k)
V_{ml}(\mathbf k)
=
V_{nl}(\mathbf k)
\varepsilon_{\mathbf kl},
\end{equation}
where $H_{nm}(\mathbf k)$ is the interpolated Wannier Hamiltonian, $\varepsilon_{\mathbf kl}$ are its band eigenvalues, and the columns of the unitary matrix $V(\mathbf k)$ are the corresponding eigenvectors. The plane-wave components of the resulting band eigenstates are then
\begin{equation}
\label{eq:inverse_oss}
u_{\mathbf kn}(\mathbf G,\sigma)
=
\sum_m
u^{\mathrm{OSS}}_{\mathbf km}(\mathbf G,\sigma)
V_{mn}(\mathbf k).
\end{equation}
The BSE transition densities can thus be constructed directly from the interpolated band eigenstates.

To evaluate a generic transition density $\rho^{\mathbf G}_{mn}(\mathbf k,\mathbf q)$, we define the final-state momentum
\begin{equation}
\mathbf k_f=\mathbf k+\mathbf q
\end{equation}
and decompose it as
\begin{equation}
\mathbf k_f
=
\widetilde{\mathbf k}_f
+
\mathbf G_{\mathbf k_f},
\end{equation}
where $\widetilde{\mathbf k}_f$ is the representative of $\mathbf k_f$ in the first BZ and $\mathbf G_{\mathbf k_f}$ is the reciprocal-lattice vector required to fold $\mathbf k_f$ into it. If $\mathbf k_f$ already lies in the first BZ, then $\widetilde{\mathbf k}_f=\mathbf k_f$ and $\mathbf G_{\mathbf k_f}=\mathbf 0$. Substituting the plane-wave expansion into Eq.~\eqref{eq:rho_periodic} gives
\begin{equation}
\label{eq:rho_pw}
\rho^{\mathbf G}_{mn}(\mathbf k,\mathbf q)
=
\sum_{\mathbf G',\sigma}
\left[
u_{\widetilde{\mathbf k}_f m}
\left(
\mathbf G'
+
\mathbf G
+
\mathbf G_{\mathbf k_f},
\sigma
\right)
\right]^*
u_{\mathbf kn}(\mathbf G',\sigma).
\end{equation}
This equation provides the transition densities entering the BSE kernels directly from the interpolated reciprocal-space Bloch states. Following the direct Wannier interpolation approach in the OSS~\cite{Volpato_2025}, the reciprocal-space interpolation scheme itself can be chosen independently of the Wannier construction. Details of the scheme adopted here are given in Appendix~\ref{subsec:interpolation_scheme}.

\subsection{Interpolation of dielectric screening}
\label{sec:epsm1_interp}

The direct interpolation of the inverse dielectric function $\epsilon^{-1}$ is inconvenient because its momentum dependence retains the rapid variation associated with the bare Coulomb interaction, particularly in the long-wavelength limit of reduced-dimensional systems. Following Ref.~\cite{Guandalini_2023}, we instead interpolate a function in which the explicit Coulomb dependence is removed. Defining
\begin{equation}
d_{\mathbf G\mathbf G'}(\mathbf q)
=
\sqrt{
v_{\mathbf G}(\mathbf q)
v_{\mathbf G'}(\mathbf q)
},
\end{equation}
and
\begin{equation}
S_{\mathbf G\mathbf G'}(\mathbf q)
=
\epsilon^{-1}_{\mathbf G\mathbf G'}(\mathbf q)
-
\delta_{\mathbf G\mathbf G'},
\end{equation}
we introduce
\begin{equation}
\label{eq:auxiliary_f}
f_{\mathbf G\mathbf G'}(\mathbf q)
=
\frac{
S_{\mathbf G\mathbf G'}(\mathbf q)
}{
d_{\mathbf G\mathbf G'}(\mathbf q)
\left[
S_{\mathbf G\mathbf G'}(\mathbf q)+1
\right]
}.
\end{equation}
All operations in Eq.~\eqref{eq:auxiliary_f} are elementwise; in particular, the scalar $1$ in the denominator is not a Kronecker delta and no matrix inversion is implied.

The function $f_{\mathbf G\mathbf G'}(\mathbf q)$ has a much smoother momentum dependence than $\epsilon^{-1}$, since the singular or rapidly varying Coulomb factors have been removed. In the absence of local-field effects, its head reduces to
\begin{equation}
f_{\mathbf 0\mathbf 0}(\mathbf q)
=
\chi^0_{\mathbf 0\mathbf 0}(\mathbf q),
\end{equation}
while in the general case it also contains the effects of local-field coupling~\cite{Guandalini_2023}. This construction is particularly advantageous for two-dimensional systems with a truncated Coulomb interaction and is also used for the non-head components in bulk systems. For three-dimensional bulk systems, however, we interpolate the head $\epsilon^{-1}_{\mathbf 0\mathbf 0}(\mathbf q)$ directly. This quantity is smooth near $\Gamma$, whereas interpolating $f_{\mathbf 0\mathbf 0}$ would require its quadratic small-$q$ behaviour to be reproduced with high accuracy in order to cancel the $1/q^2$ Coulomb divergence.

At an arbitrary target momentum $\mathbf q$, the function $f_{\mathbf G\mathbf G'}(\mathbf q)$ is interpolated using the scheme described in Appendix~\ref{subsec:interpolation_scheme}, except for the bulk head treated as described above. The Coulomb factors are then evaluated directly at the same momentum and the inverse dielectric matrix is reconstructed as
\begin{equation}
\label{eq:reconstructed_dielectric_matrix}
\epsilon^{-1}_{\mathbf G\mathbf G'}(\mathbf q)
=
\delta_{\mathbf G\mathbf G'}
+
\frac{
f_{\mathbf G\mathbf G'}(\mathbf q)
d_{\mathbf G\mathbf G'}(\mathbf q)
}{
1-
f_{\mathbf G\mathbf G'}(\mathbf q)
d_{\mathbf G\mathbf G'}(\mathbf q)
}.
\end{equation}
The screened interaction follows from
\begin{equation}
\label{eq:interpolated_W}
W_{\mathbf G\mathbf G'}(\mathbf q)
=
\sqrt{v_{\mathbf G}(\mathbf q)}
\,
\epsilon^{-1}_{\mathbf G\mathbf G'}(\mathbf q)
\,
\sqrt{v_{\mathbf G'}(\mathbf q)}.
\end{equation}
These reconstruction formulas are again understood elementwise. The bare Coulomb interaction is never interpolated, but evaluated analytically at each required momentum.

\subsection{Reconstruction of the static BSE Hamiltonian at arbitrary momentum}
\label{subsec:bse_reconstruction}

\begin{figure}[tbp]
    \centering
    \includegraphics[width=0.99\linewidth]{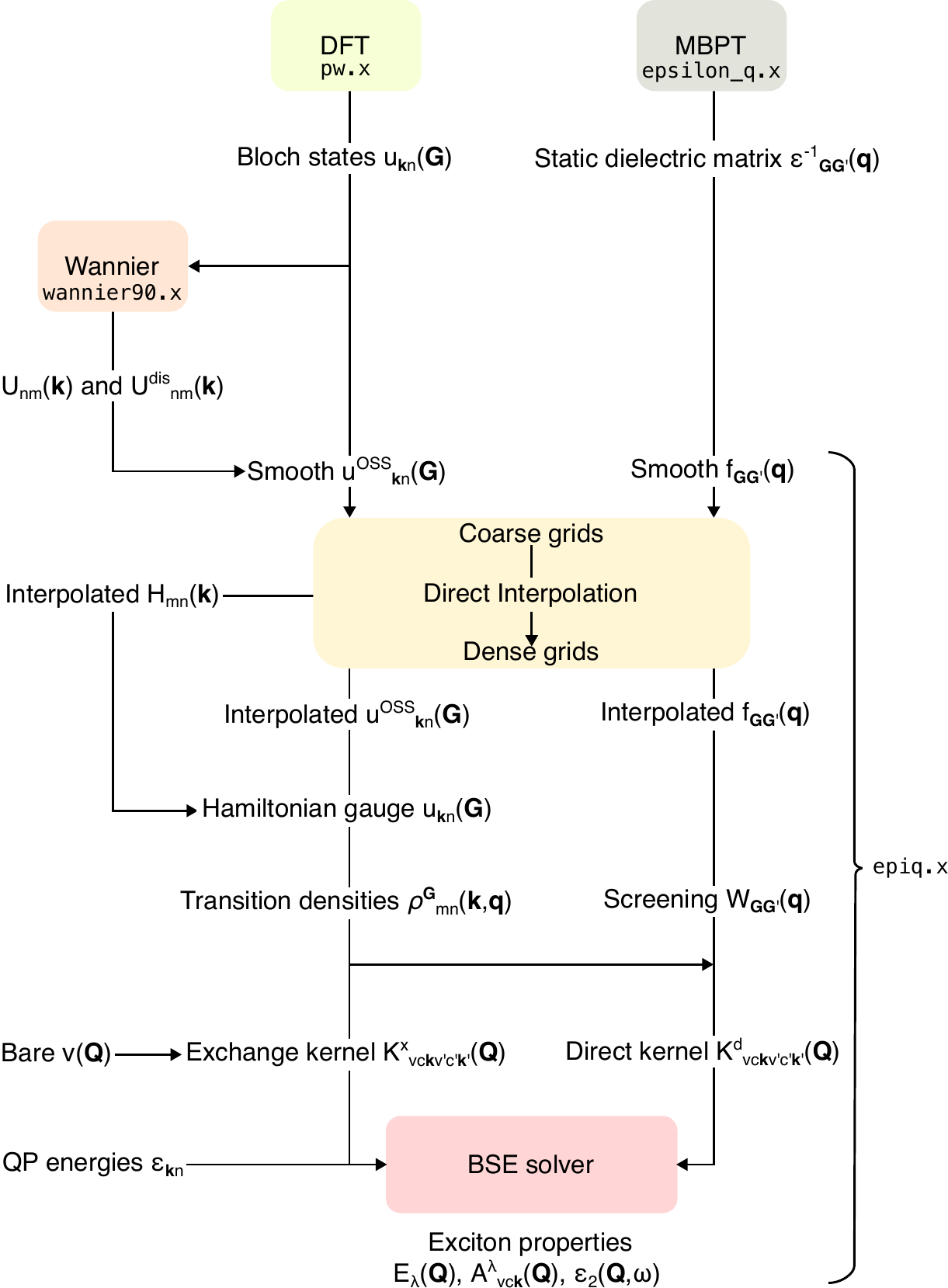}
    \caption{Schematic representation of the reciprocal-space BSE interpolation scheme. The Bloch states are obtained from density-functional theory using \texttt{pw.x} from \textsc{Quantum ESPRESSO}~\cite{Giannozzi_2009,Giannozzi_2017}, while the static dielectric matrix is computed with our \texttt{epsilon\_q.x} implementation within \textsc{Quantum ESPRESSO}. The Bloch states are transformed into the optimally smooth Wannier subspace using \textsc{Wannier90}~\cite{Pizzi_2020}, yielding the matrices $U(\mathbf k)$ and $U^{\mathrm{dis}}(\mathbf k)$. The smooth Bloch states and screening response $f_{\mathbf G\mathbf G'}(\mathbf q)$ are independently interpolated onto the target momenta. Reconstruction of the band eigenstates and screened interaction, construction of the BSE kernels, and solution of the BSE are performed within \textsc{EPIq}~\cite{Marini_2024}, yielding the exciton energies, amplitudes, and optical response.}
    \label{fig:workflow}
\end{figure}

The reconstruction of the BSE Hamiltonian from coarse first-principles inputs is summarised in Fig.~\ref{fig:workflow}. Starting from the Bloch states and static dielectric response, the electronic states are transformed into an optimally smooth Wannier representation, while the auxiliary screening response $f_{\mathbf G\mathbf G'}(\mathbf q)$ is constructed from the dielectric matrix. Both quantities are then interpolated independently at the momenta required by the target BSE calculation.

The interpolated Bloch states are rotated back to the band-eigenstate representation using Eq.~\eqref{eq:inverse_oss}, from which the transition densities are constructed through Eq.~\eqref{eq:rho_pw}. For each pair of transitions, the screened interaction is reconstructed at the exchanged momentum $\mathbf q=\mathbf k-\mathbf k'$, while the bare Coulomb interaction is evaluated directly at $\mathbf Q$. These quantities determine the direct and exchange kernels of Eqs.~\eqref{eq:direct_kernel} and~\eqref{eq:exchange_kernel}.

The screened interaction entering the direct kernel can also be averaged over the reciprocal-space cell associated with $\mathbf q$ to improve convergence with momentum sampling~\cite{Guandalini_2023}. The independent-particle contribution $H^0$ is constructed from the chosen band energies, which can be obtained from Wannier-interpolated Kohn--Sham bands, optionally supplemented by a scissor correction, or from interpolated quasiparticle energies. Combining these contributions yields the BSE Hamiltonian of Eq.~\eqref{eq:bse_hamiltonian}, which can be constructed and solved for arbitrary transition momenta $\mathbf k$, $\mathbf k'$ and transferred momentum $\mathbf Q$, independently of the underlying first-principles grids.

\subsection{Solution of the Bethe--Salpeter equation}
\label{subsec:bse_solution}

Once $H^{\mathrm{BSE}}(\mathbf Q)$ has been constructed, we employ different iterative solvers depending on the quantity of interest, avoiding explicit diagonalisation of the full electron--hole Hamiltonian.

Optical spectra are computed using the Lanczos--Haydock recursion~\cite{Lanczos_1950,Haydock_1980,Benedict_1998}. Starting from the optical transition vector, the method constructs a reduced tridiagonal representation of the BSE Hamiltonian through successive matrix--vector products. The resulting continued fraction directly yields the frequency-dependent optical response and, in particular, the imaginary part of the dielectric function, $\epsilon_2(\mathbf Q,\omega)=\mathrm{Im}\,\epsilon(\mathbf Q,\omega)$, without requiring the full excitonic eigenspectrum.

When individual excitonic states are required, we instead employ Davidson iterative diagonalisation~\cite{Davidson_1975}. The reduced subspace is iteratively refined using the residuals of the approximate eigenvectors, yielding selected low-lying exciton energies $E_\lambda(\mathbf Q)$ and amplitudes $A^\lambda_{vc\mathbf k}(\mathbf Q)$ without diagonalising the full BSE Hamiltonian.

\section{Computational details}
\label{subsec:computational_details}

Density-functional theory calculations are performed with \textsc{Quantum ESPRESSO}~\cite{Giannozzi_2009,Giannozzi_2017} within the local-density approximation (LDA)~\cite{Kohn_1965}, using optimised norm-conserving Vanderbilt pseudopotentials~\cite{Hamann_2013}. Plane-wave kinetic-energy cutoffs of 80~Ry are used for monolayer hBN and MoS$_2$, and 60~Ry for bulk GaAs. The periodic images of the monolayers are separated by 17~\AA\ of vacuum and we employ a truncated Coulomb interaction~\cite{Sohier_2017}. Ground-state charge densities are computed on $\Gamma$-centred Monkhorst--Pack meshes~\cite{Monkhorst_1976} of $18\times18\times1$ for hBN and MoS$_2$ and $12\times12\times12$ for GaAs. Spin--orbit coupling is included for MoS$_2$.

Maximally localised Wannier functions are constructed with \textsc{Wannier90}~\cite{Pizzi_2020}, retaining 10, 22, and 14 Wannier functions from manifolds of 20, 34, and 20 bands for hBN, MoS$_2$, and GaAs, respectively. Several uniform $\mathbf k$ meshes are used to assess the dependence of the interpolation accuracy on the source sampling. The reciprocal-space components of the Bloch states are extracted from the same states used for the Wannierisation through a modified version of the \texttt{wfck2r.x} utility of \textsc{Quantum ESPRESSO}. The quality of the Wannier representation is assessed in Appendix~\ref{app:wannier_quality}, where the \textit{ab initio} and Wannier-interpolated band structures are compared for the valence and conduction manifolds relevant to the BSE calculations. The initial orbital projections are also reported there.

The static RPA dielectric response used for the screening interpolation is computed with \texttt{epsilon\_q.x}, a utility implemented by us within \textsc{Quantum ESPRESSO}. The independent-particle response includes 1000 empty states and uses block-size cutoffs of 20~Ry for hBN and MoS$_2$ and 15~Ry for GaAs. The internal Brillouin-zone $\mathbf k$-point summation entering the independent-particle response, see Eq.~\eqref{eq:chi0}, is converged on $48\times48\times1$, $36\times36\times1$, and $24\times24\times24$ meshes for hBN, MoS$_2$, and GaAs, respectively. For the monolayers, the Coulomb interaction is truncated along the non-periodic direction. Several uniform $\mathbf q$ meshes are used to assess the dependence of the screening interpolation on the source sampling.

The interpolated calculations are validated against explicit \textit{ab initio} BSE calculations performed with \textsc{Yambo}~\cite{Marini_2009,Sangalli_2019}. Quasiparticle corrections are approximated by rigid scissor shifts of 2.8, 0.8, and 0.5~eV for hBN, MoS$_2$, and GaAs, respectively. In the \textit{ab initio} calculations, the scissor shifts are applied directly on the corresponding mesh, whereas in the interpolated calculations they are applied after Wannier interpolation onto the target mesh. The screened direct kernel is truncated at 10~Ry for hBN and MoS$_2$ and 8~Ry for GaAs, while a 40~Ry cutoff is used for the bare exchange term in all cases. The transition space contains the two highest valence and two lowest conduction bands for hBN, four valence and four conduction bands for MoS$_2$, and three valence and four conduction bands for GaAs. Optical spectra are evaluated with constant broadenings of 0.10, 0.075, and 0.15~eV for hBN, MoS$_2$, and GaAs, respectively.

The only external codes required by the present implementation are \textsc{Quantum ESPRESSO} for the DFT Bloch states and \textsc{Wannier90} for the Wannier representation. The dielectric response is computed with \texttt{epsilon\_q.x}, while Bloch-state and screening interpolation, construction of the BSE kernels, and solution of the BSE are performed within \textsc{EPIq}~\cite{Marini_2024}. No external many-body perturbation theory code is therefore required; \textsc{Yambo} is used only to generate independent reference calculations for validation.

\section{Benchmarks}
\subsection{Interpolation accuracy of Bloch states and dielectric screening}
\label{subsec:interpolation_accuracy}

We first benchmark the Bloch-state and dielectric-screening interpolation separately before assessing their combined accuracy at the BSE level. Monolayer MoS$_2$ is used as a representative case. Starting from a $12\times12\times1$ source mesh, both quantities are interpolated on a $36\times36\times1$ target grid and compared point by point with explicit first-principles calculations. This directly tests the interpolation at momenta absent from the source grid.

\begin{figure}[tbp]
    \centering
    \includegraphics[width=0.99\linewidth]{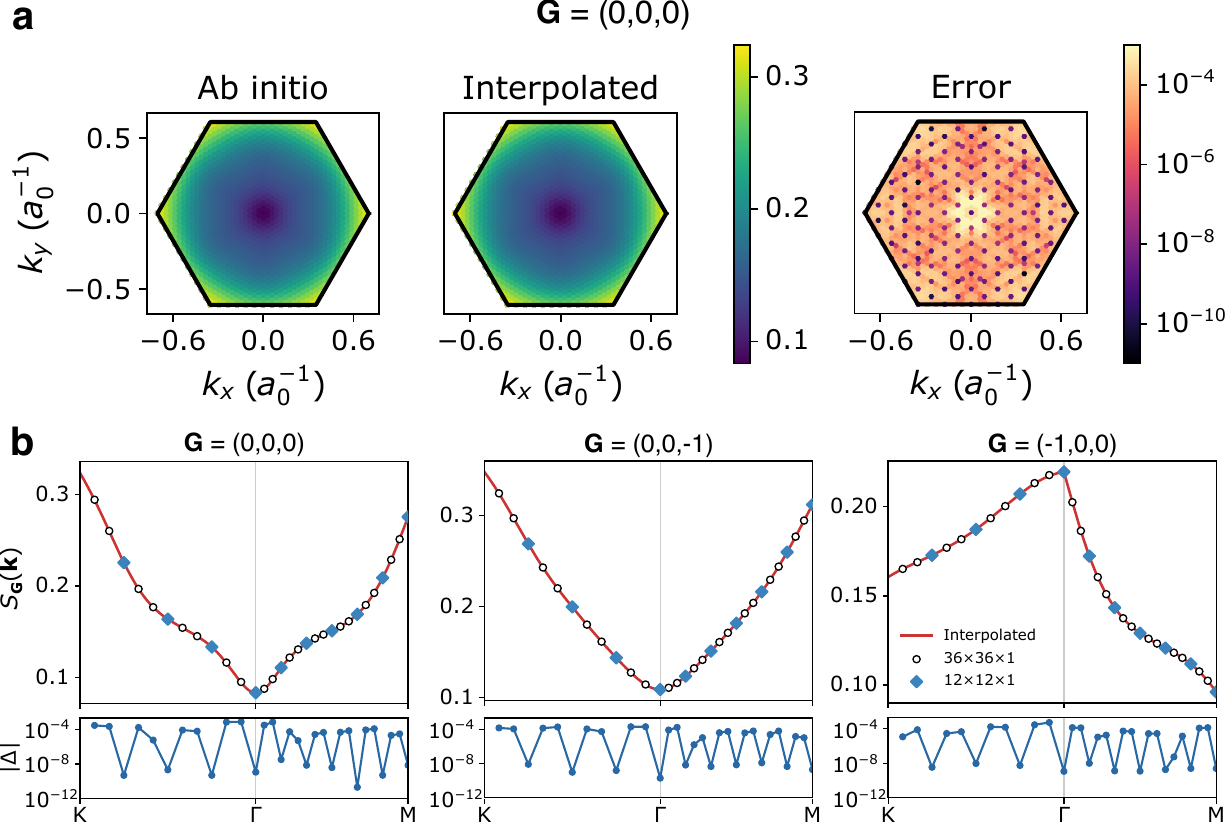}
    \caption{Interpolation benchmark for the Bloch-state plane-wave components of monolayer MoS$_2$, expressed through the unitary-invariant quantity $S_{\mathbf G}(\mathbf k)$ defined in Eq.~\eqref{eq:S_G}. (a) $S_{\mathbf G}(\mathbf k)$ for $\mathbf G=(0,0,0)$ from the explicit $36\times36\times1$ calculation (left), interpolation from a $12\times12\times1$ source mesh (centre), and absolute interpolation error (right). (b) $S_{\mathbf G}(\mathbf k)$ along the $K$--$\Gamma$--$M$ path for three representative reciprocal-lattice vectors. Red lines denote the interpolated values, open circles the explicit $36\times36\times1$ reference, and blue diamonds the $12\times12\times1$ source points. The lower panels show the absolute interpolation error at the reference momenta.}
    \label{fig:wfc_interpolation}
\end{figure}

Figure~\ref{fig:wfc_interpolation} assesses the Bloch-state interpolation through the quantity
\begin{equation}
\label{eq:S_G}
S_{\mathbf G}(\mathbf k)
=
\sum_{n\sigma}
\left|
u_{\mathbf kn}(\mathbf G,\sigma)
\right|^2,
\end{equation}
where the sum runs over all bands spanning the Wannier subspace and over spin. Since $S_{\mathbf G}(\mathbf k)$ is invariant under unitary rotations within the subspace, it probes the plane-wave content of the interpolated Bloch manifold independently of the representation of the individual eigenstates.

For $\mathbf G=(0,0,0)$, the interpolated Brillouin-zone map is essentially indistinguishable from the explicit calculation, with a maximum relative error below $1\%$ over the full target grid, as shown in Fig.~\ref{fig:wfc_interpolation}(a). Similarly, along the $K$--$\Gamma$--$M$ path, the interpolated curves closely reproduce the reference values for the representative reciprocal-lattice vectors shown in Fig.~\ref{fig:wfc_interpolation}(b). The interpolation therefore accurately reproduces the plane-wave content of the Bloch subspace throughout the Brillouin zone. 

\begin{figure}[tbp]
    \centering
    \includegraphics[width=0.99\linewidth]{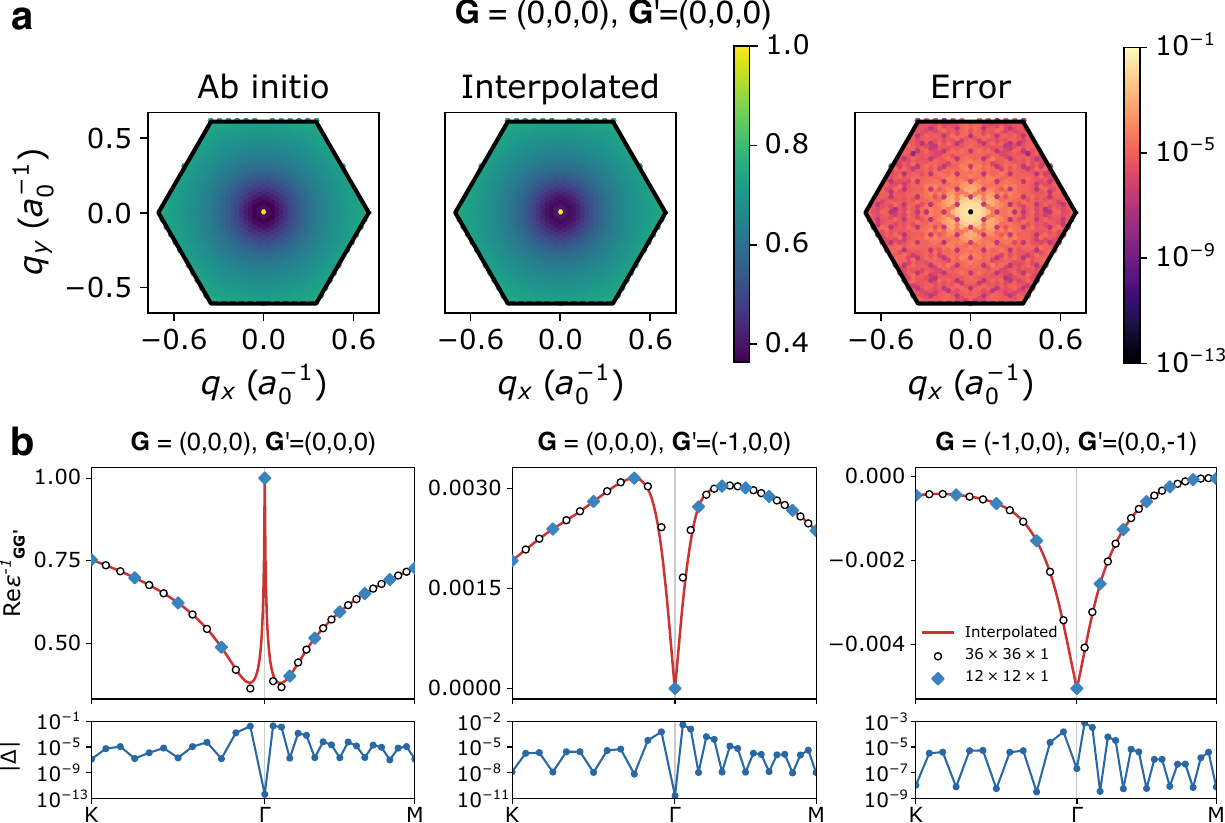}
    \caption{Interpolation benchmark for the static dielectric response of monolayer MoS$_2$. (a) Real part of the head element $\epsilon^{-1}_{\mathbf G\mathbf G'}(\mathbf q)$, with $\mathbf G=\mathbf G'=(0,0,0)$, from the explicit $36\times36\times1$ calculation (left), reconstructed from the screening response interpolated from a $12\times12\times1$ source mesh (centre), and modulus of the interpolation error (right). (b) Representative elements of $\mathrm{Re}\,\epsilon^{-1}_{\mathbf G\mathbf G'}(\mathbf q)$ along the $K$--$\Gamma$--$M$ path. Red lines denote the reconstructed values, open circles the explicit $36\times36\times1$ reference, and blue diamonds the $12\times12\times1$ source points. The lower panels show the modulus of the complex interpolation error at the reference momenta.}
    \label{fig:screening_interpolation}
\end{figure}

The dielectric response provides a more stringent test because of its rapid variation in the long-wavelength region. Figure~\ref{fig:screening_interpolation} compares the reconstructed and explicit inverse dielectric matrix $\epsilon^{-1}_{\mathbf G\mathbf G'}(\mathbf q)$. The quantity directly interpolated is the screening response $f_{\mathbf G\mathbf G'}(\mathbf q)$ introduced in Eq.~\eqref{eq:auxiliary_f}, from which $\epsilon^{-1}_{\mathbf G\mathbf G'}(\mathbf q)$ is reconstructed using Eq.~\eqref{eq:reconstructed_dielectric_matrix}. The head element, $\mathbf G=\mathbf G'=(0,0,0)$, is accurately reproduced over most of the Brillouin zone, as shown in Fig.~\ref{fig:screening_interpolation}(a), with similarly good agreement for representative diagonal and off-diagonal elements along $K$--$\Gamma$--$M$ in Fig.~\ref{fig:screening_interpolation}(b).

The largest deviations occur for the head element close to $\Gamma$, where the dielectric response varies on a momentum scale only sparsely resolved by the $12\times12\times1$ source mesh. Even there, the maximum relative error remains around $5\%$ and decreases rapidly away from the long-wavelength limit. The screening reconstruction is therefore already accurate for this deliberately coarse source grid, while its convergence remains sensitive to the source sampling. Refining this mesh progressively resolves the sharp variation around $\Gamma$ and systematically improves the accuracy, as examined below.

Together, Figs.~\ref{fig:wfc_interpolation} and~\ref{fig:screening_interpolation} show that Bloch states and dielectric screening can be accurately interpolated well beyond the original first-principles sampling, including at momenta absent from the source grids. This enables the BSE sampling to be decoupled from the underlying \textit{ab initio} calculation. We next assess the combined accuracy at the level of the full electron--hole Hamiltonian.

\subsection{Reconstructed BSE optical spectra}
\label{subsec:bse_spectra}

\begin{figure*}[t]
    \centering
    \includegraphics[width=0.99\linewidth]{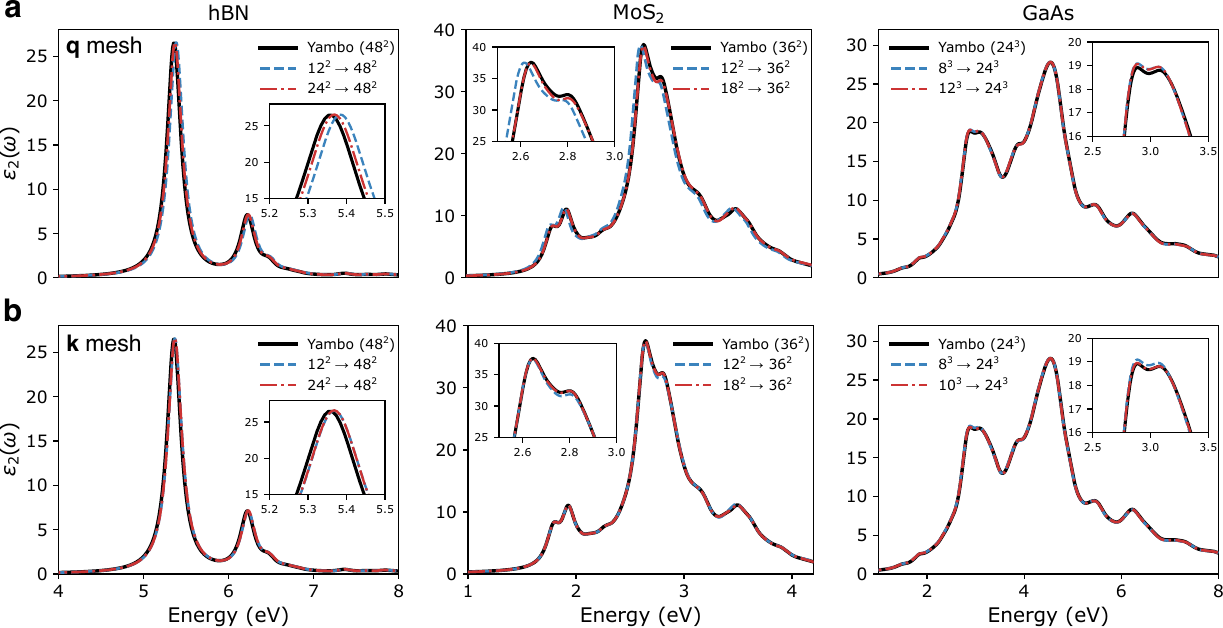}
    \caption{BSE optical spectra obtained from the reciprocal-space interpolation scheme compared with explicit \textit{ab initio} calculations for monolayer hBN and MoS$_2$ and bulk GaAs. Black solid lines show the explicit \textsc{Yambo} reference spectra, calculated on $48\times48\times1$, $36\times36\times1$, and $24\times24\times24$ meshes, respectively. Blue dashed and red dash-dotted lines show reconstructed spectra; the legend indicates the source and target BSE meshes as source $\rightarrow$ target. (a) Convergence with respect to the source $\mathbf q$ mesh used for the dielectric screening, with the Bloch-state source mesh fixed at the coarser sampling. (b) Convergence with respect to the source $\mathbf k$ mesh used for the Bloch states, with the dielectric screening fixed at the finer $\mathbf q$ sampling. Insets magnify representative excitonic features where the residual differences are most visible.}
    \label{fig:bse_spectra}
\end{figure*}

Having benchmarked the Bloch states and dielectric screening separately, we now assess their combined accuracy at the level of the BSE optical response. Figure~\ref{fig:bse_spectra} compares spectra obtained from the reconstructed BSE Hamiltonian with explicit \textit{ab initio} calculations for monolayer hBN and MoS$_2$ and bulk GaAs. The reference calculations use $48\times48\times1$, $36\times36\times1$, and $24\times24\times24$ meshes for hBN, MoS$_2$, and GaAs, respectively, and the reconstructed spectra are evaluated on the same target grids. The Bloch states and dielectric screening entering the reconstructed Hamiltonian are instead obtained from substantially coarser source meshes.

To separate the sources of interpolation error, we first examine convergence with respect to the source $\mathbf q$ mesh used for the dielectric screening, as shown in Fig.~\ref{fig:bse_spectra}(a), while keeping the Bloch-state mesh fixed at the coarser sampling. We then refine the Bloch-state source $\mathbf k$ mesh in Fig.~\ref{fig:bse_spectra}(b), with the screening fixed at the finer $\mathbf q$ sampling.

For all three materials, the reconstructed spectra closely reproduce the explicit reference, including excitonic line shapes, peak positions, and oscillator strengths, already from the coarser source meshes. The remaining differences are mainly small energy shifts and localised changes in spectral intensity, highlighted by the insets, and decrease systematically upon refinement.

Convergence with respect to the screening mesh in Fig.~\ref{fig:bse_spectra}(a) is the more demanding of the two, consistent with the benchmark of Fig.~\ref{fig:screening_interpolation}. In hBN, refining the source $\mathbf q$ mesh from $12\times12\times1$ to $24\times24\times1$ substantially reduces the shift of the dominant excitonic peak. The improvement is even stronger for MoS$_2$, where the main excitonic structure becomes essentially indistinguishable from the explicit result upon refinement from $12\times12\times1$ to $18\times18\times1$. For GaAs, the spectrum is already well reproduced from the $8\times8\times8$ source mesh, with only minor changes upon refinement to $12\times12\times12$. These trends reflect the need to resolve the momentum dependence of the dielectric response on the source grid, particularly in low-dimensional systems.

Figure~\ref{fig:bse_spectra}(b) shows the convergence with respect to the Bloch-state $\mathbf k$ mesh, with the screening fixed at the finer sampling. For hBN, the two reconstructed spectra are nearly indistinguishable, indicating that the remaining discrepancy is dominated by the screening. In MoS$_2$ and GaAs, refining the $\mathbf k$ mesh mainly corrects small localised intensity differences, while leaving the overall line shape and peak positions essentially unchanged. This is consistent with the sub-percent errors found in the unitary-invariant Bloch-state benchmark of Fig.~\ref{fig:wfc_interpolation} and confirms that the interpolated plane-wave components accurately reproduce the transition densities entering the electron--hole kernels.

Because the residual differences are dominated by small energy shifts and localised redistributions of spectral weight, a pointwise relative error is not a particularly meaningful measure of spectral agreement. We therefore quantify the convergence using the first Wasserstein distance between the normalised spectra. Defining the cumulative spectral weight
\begin{equation}
F(\omega)
=
\dfrac{\displaystyle\int_{-\infty}^{\omega}
d\omega'\,\epsilon_2(\omega')}{\displaystyle\int_{-\infty}^{+\infty} d\omega'\,\epsilon_2(\omega')},
\end{equation}
the spectral distance is
\begin{equation}
\label{eq:spectral_wasserstein}
W_1
=
\int d\omega\,
\left|
F(\omega)-F_{\mathrm{ref}}(\omega)
\right|.
\end{equation}
For two identical line shapes rigidly displaced by $\Delta E$, this gives simply $W_1=|\Delta E|$.

The values in Table~\ref{tab:spectral_wasserstein} quantify the convergence visible in Fig.~\ref{fig:bse_spectra}. Refining the screening mesh reduces $W_1$ from 24.1 to 9.5~meV in hBN and from 23.1 to 1.7~meV in MoS$_2$, while in GaAs it decreases from 8.9 to 8.0~meV. Subsequent refinement of the Bloch-state mesh reduces the distance further to 9.0, 0.3, and 4.9~meV for hBN, MoS$_2$, and GaAs, respectively. The relative importance of the two interpolated quantities varies among the materials, but in all cases the final spectra closely reproduce the explicit reference.

These results provide the central validation of the reconstruction scheme. Starting from coarse Bloch-state and screening meshes, the method reproduces the optical response of much denser explicit calculations without interpolating the final spectrum or excitonic eigenvalues. Instead, the BSE Hamiltonian itself is reconstructed from the interpolated Bloch states and screened interaction.

\begin{table}[tbp]
\begin{ruledtabular}
\begin{tabular}{lccc}
Material & Mesh type & Source-mesh refinement & $W_1$ (meV) \\
\hline
hBN
& $\mathbf q$
& $12\times12\times1 \rightarrow 24\times24\times1$
& 24.1 $\rightarrow$ 9.5 \\
& $\mathbf k$
& $12\times12\times1 \rightarrow 24\times24\times1$
& 9.5 $\rightarrow$ 9.0 \\[2pt]

MoS$_2$
& $\mathbf q$
& $12\times12\times1 \rightarrow 18\times18\times1$
& 23.1 $\rightarrow$ 1.7 \\
& $\mathbf k$
& $12\times12\times1 \rightarrow 18\times18\times1$
& 1.7 $\rightarrow$ 0.3 \\[2pt]

GaAs
& $\mathbf q$
& $8\times8\times8 \rightarrow 12\times12\times12$
& 8.9 $\rightarrow$ 8.0 \\
& $\mathbf k$
& $8\times8\times8 \rightarrow 10\times10\times10$
& 8.0 $\rightarrow$ 4.9 \\
\end{tabular}
\end{ruledtabular}
\caption{Convergence of the reconstructed BSE optical spectra with respect to the source $\mathbf q$ mesh used for the dielectric screening and the source $\mathbf k$ mesh used for the Bloch states. During the $\mathbf q$-mesh convergence, the Bloch-state source mesh is kept fixed at the coarser sampling; during the subsequent $\mathbf k$-mesh convergence, the screening is kept fixed at the finer sampling. The arrows indicate refinement of the source mesh, while the target BSE meshes are those specified in Fig.~\ref{fig:bse_spectra}. $W_1$ is the first Wasserstein distance from the explicit \textit{ab initio} BSE spectrum, defined in Eq.~\eqref{eq:spectral_wasserstein}.}
\label{tab:spectral_wasserstein}
\end{table}

\section{Results}
\subsection{Computational cost and scaling}
\label{subsec:computational_cost}

\begin{figure}[tbp]
    \centering
    \includegraphics[width=0.99\linewidth]{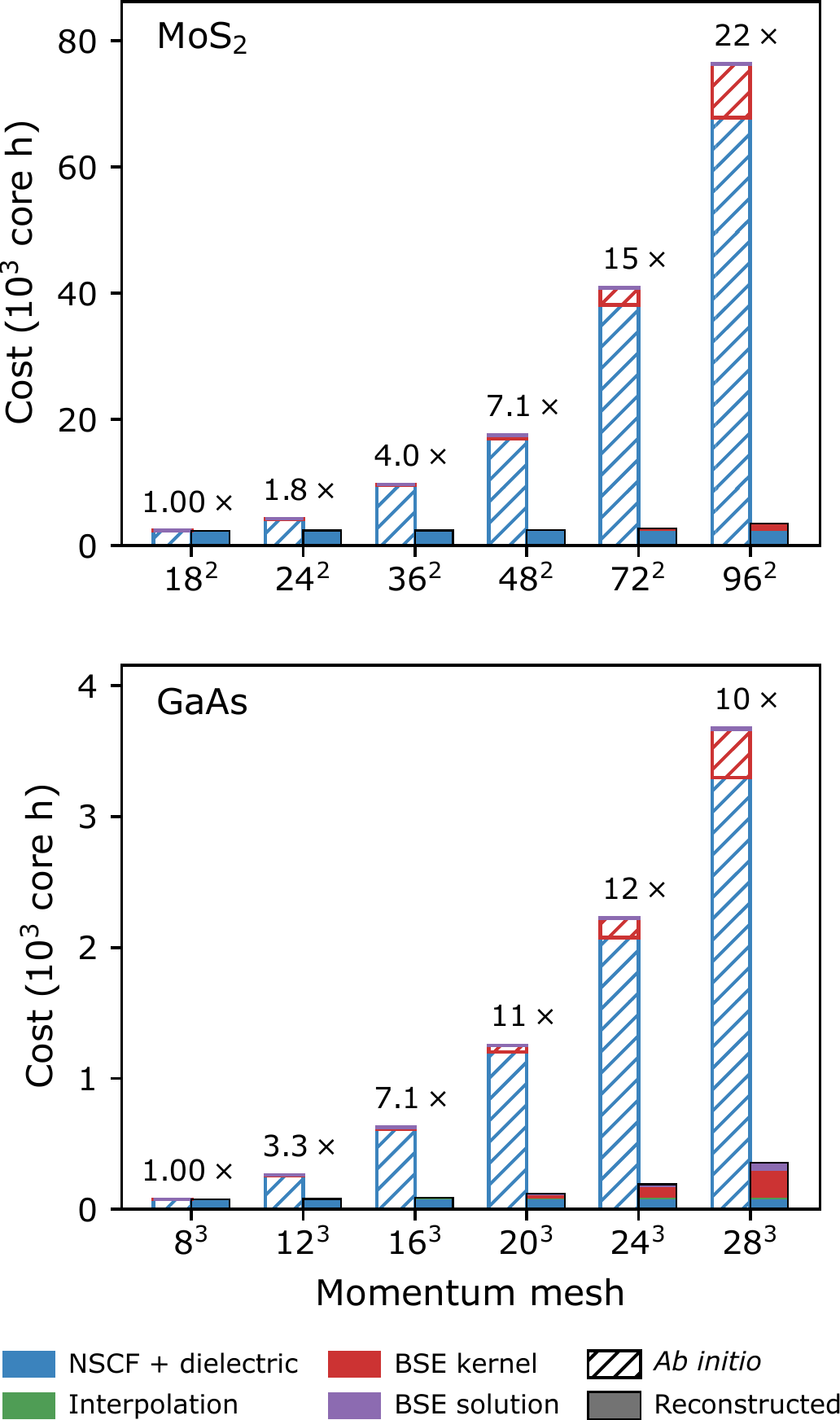}
    \caption{Computational cost of explicit \textit{ab initio} and reconstructed BSE calculations as a function of the target momentum mesh for monolayer MoS$_2$ (top) and bulk GaAs (bottom). The cost is reported in core hours and decomposed into the first-principles calculation of the Bloch states and dielectric response, interpolation to the target momenta, construction of the transition densities and BSE kernel, and iterative solution of the BSE. Hatched bars denote explicit calculations and filled bars reconstructed calculations. In the latter, the Bloch states and dielectric response are computed only on fixed source meshes, $18^2$ for MoS$_2$ and $8^3$ for GaAs, and subsequently interpolated to the target sampling. Numbers above the bars give the ratio between the total explicit and reconstructed computational costs.}
    \label{fig:computational_cost}
\end{figure}

The accuracy demonstrated above is accompanied by a substantial reduction in the cost of dense BSE calculations. Figure~\ref{fig:computational_cost} compares the cost of obtaining BSE optical spectra for monolayer MoS$_2$ and bulk GaAs from explicit calculations on the target mesh and from BSE Hamiltonians reconstructed using Bloch states and screening interpolated from coarser source meshes. Timings are reported in core hours and decomposed into the first-principles calculation of Bloch states and dielectric response, interpolation to the target momenta, construction of the transition densities and BSE kernel, and iterative solution of the BSE. Reference timings are measured on $36\times36\times1$ for MoS$_2$ and $24\times24\times24$ for GaAs; the remaining costs are obtained from the scaling relations discussed below.

The first-principles contribution includes the non-self-consistent-field (NSCF) calculation of the Bloch states and the dielectric response. In the explicit calculations, the NSCF $\mathbf k$ mesh is refined together with the $\mathbf q$ mesh required for screening. For fixed band and plane-wave cutoffs, the NSCF cost scales approximately with the number of Bloch momenta, $N_k$. The independent-particle polarisability of Eq.~\eqref{eq:chi0} must instead be evaluated for every momentum transfer $\mathbf q$, involving an internal $\mathbf k$-point summation. In our cost estimates, this internal sampling is kept fixed at $36\times36\times1$ for MoS$_2$ and $24\times24\times24$ for GaAs, as specified in Sec.~\ref{subsec:computational_details}. The dielectric-response cost therefore scales linearly with the number of sampled momentum transfers, $N_q$, giving
\begin{equation}
T_{\mathrm{FP}}
\propto N_k+N_q.
\end{equation}
This is a conservative assumption: in conventional implementations such as \textsc{Yambo}, the internal $\mathbf k$ mesh entering $\chi^0$ is typically refined together with the BSE sampling, increasing the cost of the explicit screening calculation. By contrast, in the reconstructed calculations, the Bloch states and dielectric response are computed only on the fixed $18\times18\times1$ and $8\times8\times8$ source meshes for MoS$_2$ and GaAs, respectively, so their cost remains constant as the target mesh is refined. The speedups reported below therefore represent conservative estimates of the computational advantage of interpolation.

The interpolation cost also grows with the target sampling. Bloch states are reconstructed once at each required $\mathbf k$ point and stored, while the dielectric response is reconstructed independently at each required $\mathbf q$. Hence,
\begin{equation}
T_{\mathrm{interp}}
\propto N_k+N_q.
\end{equation}
The subsequent BSE-kernel construction has the same momentum scaling in the explicit and reconstructed calculations. On the uniform meshes considered here, the transition densities and interaction matrix elements must be evaluated over all $(\mathbf k,\mathbf q)$ pairs. For fixed band and plane-wave cutoffs, this gives
\begin{equation}
T_{\mathrm{kernel}}
\propto N_kN_q.
\end{equation}
The iterative BSE solution is dominated by Hamiltonian applications over the electron--hole transition space and, for fixed numbers of valence and conduction bands, scales approximately as
\begin{equation}
T_{\mathrm{BSE}}
\propto N_k^2.
\end{equation}
For the uniform samplings considered here, $N_q$ grows together with $N_k$, so both kernel construction and BSE solution scale quadratically with the target mesh size.

The resulting trend is shown in Fig.~\ref{fig:computational_cost}. At the source meshes, the explicit and reconstructed calculations have comparable costs because they require the same coarse first-principles ingredients. The advantage grows rapidly as the target mesh is refined. For MoS$_2$, the speed-up reaches approximately $22\times$ at $96\times96\times1$. In the explicit calculation, the increasing cost is dominated by the dielectric response on the dense $\mathbf q$ mesh together with the NSCF calculation of the corresponding Bloch states. In the reconstructed calculation, both first-principles contributions remain fixed at the source-mesh cost, while the interpolation overhead remains comparatively small.

For bulk GaAs, the speed-up reaches about $10\times$ at $28\times28\times28$. At dense samplings, the gain becomes smaller because construction of the transition densities and BSE kernel accounts for an increasing fraction of the reconstructed calculation. This stage is common to both approaches and retains the same $N_kN_q$ scaling. The interpolation scheme therefore does not alter the asymptotic cost of assembling the BSE Hamiltonian, but removes the need to recompute Bloch states and dielectric screening from first principles on each dense target mesh.

The BSE Hamiltonian can therefore be constructed on dense target meshes from coarse first-principles inputs at substantially lower cost than the corresponding explicit \textit{ab initio} calculation. Together with the spectral agreement of Fig.~\ref{fig:bse_spectra}, this makes flexible momentum sampling practical for the applications discussed below.

\subsection{Finite-momentum excitons in the long-wavelength limit}
\label{subsec:finite_q_exciton}

\begin{figure}[tbp]
    \centering
    \includegraphics[width=0.99\linewidth]{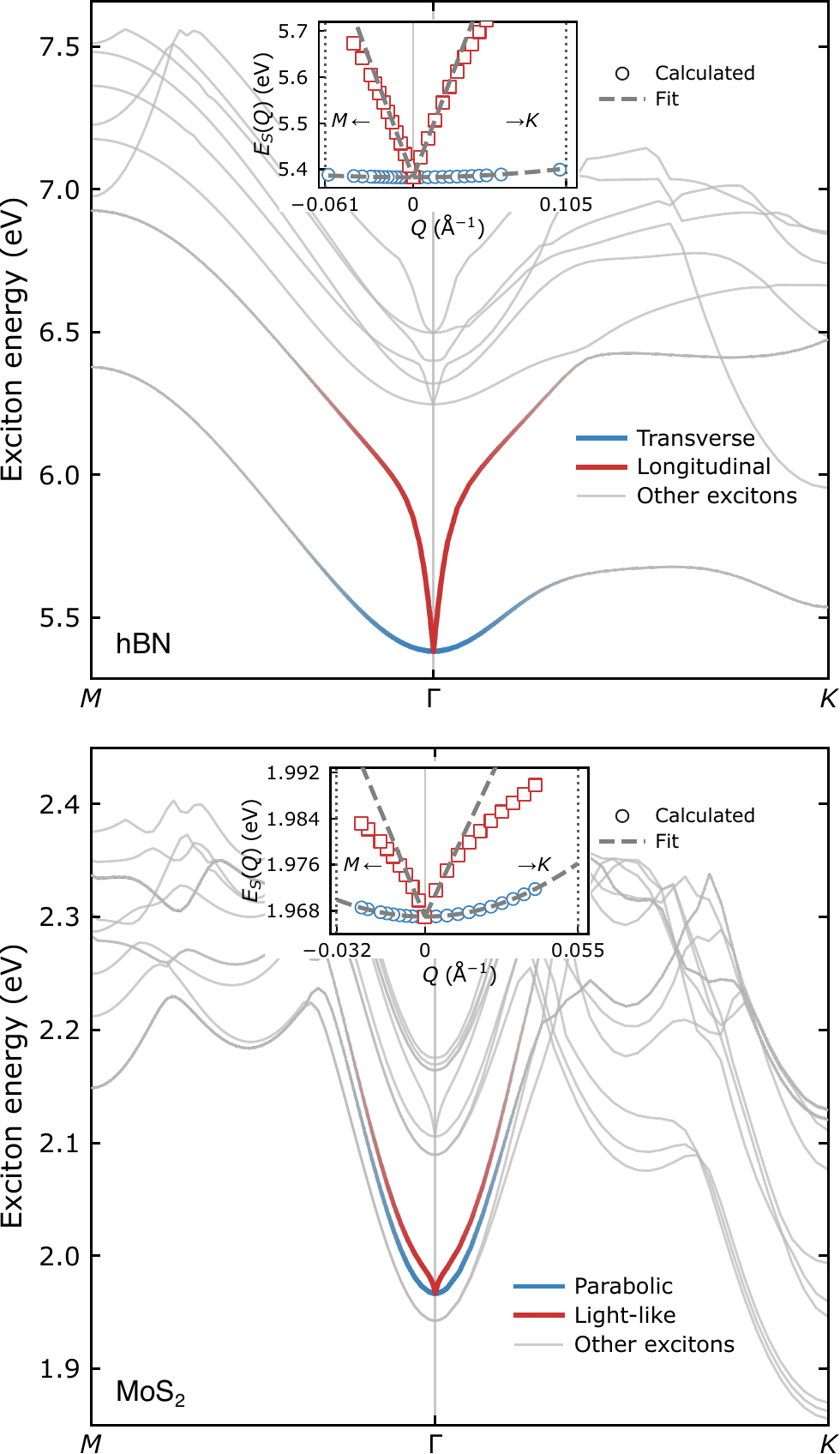}
    \caption{Finite-momentum exciton dispersion of monolayer hBN (top) and MoS$_2$ (bottom) along $M$--$\Gamma$--$K$. Near $\Gamma$, the lowest hBN doublet separates into transverse (blue) and longitudinal (red) branches, while the bright MoS$_2$ $A$-exciton doublet separates into parabolic (blue) and light-like (red) branches. Other excitons are shown in grey. Insets resolve the long-wavelength region; open symbols denote calculated energies and dashed lines the corresponding fits. Dotted vertical lines mark the closest non-zero transferred momenta to $\Gamma$ accessible on uniform $48\times48\times1$ (hBN) and $72\times72\times1$ (MoS$_2$) meshes, highlighting the finer resolution enabled by the interpolation.}
    \label{fig:exciton_dispersion}
\end{figure}

A key advantage of the present interpolation scheme is that the transferred momentum $\mathbf Q$ can be chosen independently of the underlying first-principles grids. Figure~\ref{fig:exciton_dispersion} shows the exciton dispersions of monolayer hBN and MoS$_2$ along the full $M$--$\Gamma$--$K$ path, with a finely resolved region around $\Gamma$. The exciton energies themselves are not interpolated: for each target $\mathbf Q$, the required Bloch states and screened interaction are reconstructed and a new BSE Hamiltonian is solved.

In hBN, the lowest exciton doublet at $\Gamma$ separates into a weakly dispersive transverse branch (T) and a longitudinal branch (L) with a nonanalytic dispersion induced by long-range electron--hole exchange~\cite{Andreani_1990,Qiu_2015,Cudazzo_2016}. In the long-wavelength limit,
\begin{align}
E_{\mathrm T}(Q)
&=
E_0+\frac{\hbar^2Q^2}{2M_{\mathrm T}},
\\
E_{\mathrm L}(Q)-E_{\mathrm T}(Q)
&=
2A|Q|,
\end{align}
where $E_0$ is the common energy at $\Gamma$ and $M_{\mathrm T}$ is the transverse exciton mass. Taking advantage of the arbitrary $\mathbf Q$ sampling, we fit only the first few non-zero momenta closest to $\Gamma$, where this limit applies. We obtain $A=4.26$ and $4.14$~eV\,\AA{} along $\Gamma$--$M$ and $\Gamma$--$K$, respectively, indicating a nearly isotropic linear splitting. The corresponding exchange velocities $v=2A/\hbar$ are $1.29\times10^6$ and $1.26\times10^6$~m\,s$^{-1}$, while the fitted transverse masses are $M_{\mathrm T}=3.46\,m_e$ and $2.20\,m_e$. The pronounced linear splitting is consistent with the strong long-range electron--hole exchange reported for monolayer hBN~\cite{Liu_2026,Hong_2026}.

In MoS$_2$, spin--orbit coupling separates the low-energy excitons into unlike-spin dark states and a like-spin bright $A$-exciton doublet. Exchange vanishes in the unlike-spin channel, while at finite $Q$ it splits the bright doublet into a parabolic ($-$) and a light-like ($+$) branch~\cite{Qiu_2015}. Their long-wavelength dispersion is
\begin{align}
E_{-}(Q)
&=
E_0+\frac{\hbar^2Q^2}{2M_{-}},
\\
E_{+}(Q)-E_{-}(Q)
&=
2A|Q|.
\end{align}
Using the two closest non-zero momenta, we obtain $A=0.57$ and $0.51$~eV\,\AA{} along $\Gamma$--$M$ and $\Gamma$--$K$, corresponding to velocities of $1.75\times10^5$ and $1.55\times10^5$~m\,s$^{-1}$. The fitted linear coefficients are therefore comparable along the two directions. The corresponding parabolic masses are $M_{-}=1.32\,m_e$ and $1.26\,m_e$. Quantitative agreement with previous literature is not expected from the present scissor-corrected LDA electronic structure, but the weaker light-like splitting relative to hBN is consistent with the expected exchange-driven behaviour.

These fits probe transferred momenta well below the resolution of the corresponding uniform meshes. The dotted lines in the insets of Fig.~\ref{fig:exciton_dispersion} mark the closest non-zero momenta to $\Gamma$ accessible on those meshes. For hBN, their moduli along $\Gamma$--$M$ and $\Gamma$--$K$ are $|\mathbf Q|=0.0606$ and $0.1049$~\AA$^{-1}$ on a $48\times48\times1$ mesh, whereas the fits extend only to $|\mathbf Q|=0.0087$ and $0.0151$~\AA$^{-1}$. For MoS$_2$, the corresponding values on a $72\times72\times1$ mesh are $0.0319$ and $0.0552$~\AA$^{-1}$, compared with $0.0046$ and $0.0080$~\AA$^{-1}$ in the fits. In both materials, the sampled momenta are therefore about seven times smaller in modulus. This resolution is essential because the linear form already begins to deviate from the calculated dispersion very close to $\Gamma$. Reaching the same regime with uniform grids would require substantially denser first-principles calculations of both Bloch states and screening. Recent work has accessed it through an approximate long-wavelength treatment~\cite{Guandalini_2026}; here, it is obtained directly by constructing and solving the BSE Hamiltonian at each $\mathbf Q$, without introducing an additional long-wavelength approximation.

This illustrates the main advantage of decoupling the BSE sampling from the source grids: transferred momenta can be placed directly where the exciton dispersion varies most rapidly, giving access to the long-wavelength regime without uniformly refining the Brillouin zone.

\subsection{Convergence of exciton binding energies with locally refined sampling}
\label{subsec:nonuniform_sampling}

The freedom to choose the transition-momentum sampling independently of the underlying first-principles grids allows BSE points to be concentrated where they are most relevant. This is particularly useful for weakly bound Wannier--Mott excitons~\cite{Sham_1966}, whose amplitudes are localised in reciprocal space around the band extrema. Uniform meshes can therefore converge slowly, whereas local refinement provides the required momentum resolution with far fewer points~\cite{Rohlfing_1998,Fuchs_2008,Alvertis_2023}.

We illustrate this for bulk GaAs, where the lowest exciton wave functions  at $\mathbf Q=\mathbf 0$ are concentrated around $\Gamma$. Alongside uniform meshes, we consider locally refined samplings obtained by retaining points within a sphere of radius $0.04$~\AA$^{-1}$ around $\Gamma$ from progressively denser uniform meshes. The parent uniform mesh determines the local momentum resolution, while only the retained points enter the BSE transition space. In both cases, the Bloch states and screened interaction are reconstructed from the same coarse first-principles inputs. The binding energy of exciton $\lambda$ is defined as
\begin{equation}
E_b^\lambda
=
E_g-E_\lambda(\mathbf Q=\mathbf 0),
\end{equation}
where $E_g$ is the single-particle gap entering the BSE.

\begin{figure}[tbp]
    \centering
    \includegraphics[width=0.99\linewidth]{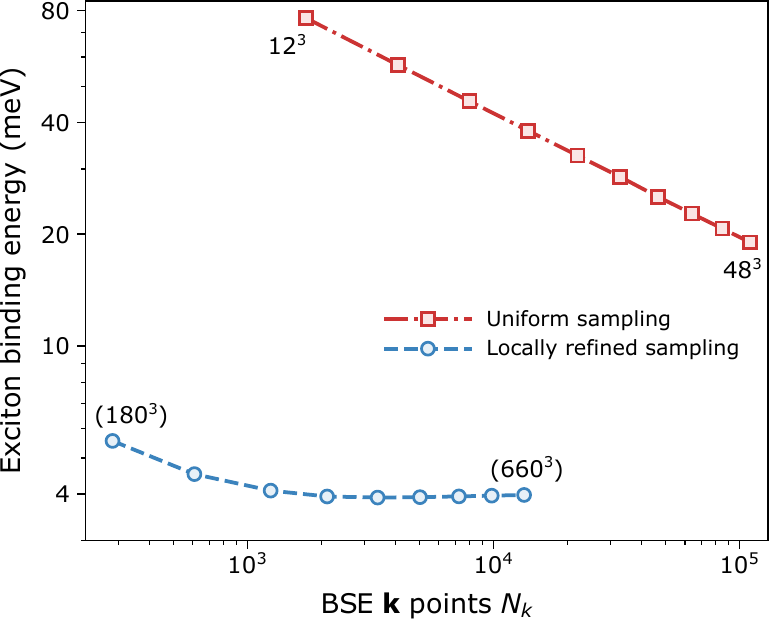}
    \caption{Convergence of the binding energy of the threefold-degenerate lowest exciton in bulk GaAs with the number of BSE $\mathbf k$ points. Red squares denote uniform sampling and blue circles locally refined sampling around $\Gamma$. Labels on the uniform series indicate the mesh dimensions, while those in parentheses indicate the uniform meshes required to achieve comparable momentum resolution near $\Gamma$.}
    \label{fig:binding_convergence}
\end{figure}

Figure~\ref{fig:binding_convergence} shows the convergence of the binding energy of the threefold-degenerate lowest exciton. Uniform sampling remains far from convergence even on a $48\times48\times48$ mesh, containing more than $10^5$ transition momenta. In contrast, local refinement yields a converged binding energy of approximately $4.0$~meV with only a few thousand points. This agrees closely with experimental values of $3.8$--$4.2$~meV~\cite{Bimberg_1972,Sell_1972,Nam_1976} and the Wannier--Mott model prediction of $4.1$~meV~\cite{Dandrea_1982}.

Earlier first-principles $GW$--BSE calculations by Rohlfing and Louie obtained a binding energy of $4.0$~meV using locally concentrated momentum sampling~\cite{Rohlfing_1998}. Our result agrees closely with this benchmark, while the present approach constructs the BSE kernels from independently interpolated Bloch states and reconstructed dielectric screening.

These results demonstrate that accurate exciton binding energies can be obtained with substantially fewer transition momenta by refining the BSE sampling locally, without increasing the underlying first-principles grids.

\section{Conclusions}

We have developed a reciprocal-space framework for reconstructing the static Bethe--Salpeter Hamiltonian at arbitrary crystal momenta from coarse first-principles grids. Bloch states are interpolated within an optimally smooth Wannier subspace, while the screened Coulomb interaction is reconstructed from an independently interpolated dielectric response. The resulting BSE Hamiltonian can be evaluated on momentum grids chosen independently of the underlying first-principles sampling.

Although we use static screening within the random-phase approximation here, the interpolation requires only the microscopic inverse dielectric matrix on a coarse $\mathbf q$ mesh, independently of the approximation used to calculate it. The dielectric response can therefore be obtained from different many-body implementations, such as \textsc{Yambo}~\cite{Marini_2009,Sangalli_2019}, \textsc{BerkeleyGW}~\cite{Deslippe_2012}, or \textsc{ABINIT}~\cite{Gonze_2020}. Likewise, the single-particle energies can be obtained from various electronic-structure approaches, including semilocal and hybrid functionals, Koopmans-compliant functionals~\cite{Nguyen_2018,Elliot_2019}, DFT+$U$, and $GW$ quasiparticle calculations. The use of a Wannier representation is not a major practical limitation, as automated procedures now allow Wannier Hamiltonians to be constructed with limited material-specific tuning, as demonstrated in high-throughput studies across broad classes of materials~\cite{Vitale_2020,Qiao_2023}. These features make the method applicable to a broad range of materials, levels of theory, and electronic-structure codes.

Benchmarks against explicit first-principles calculations show excellent agreement for the interpolated Bloch-state plane-wave components and reconstructed dielectric screening. The resulting BSE optical spectra also closely reproduce the reference calculations for monolayer hBN, MoS$_2$, and bulk GaAs. Our cost analysis estimates speedups of up to $22\times$ for MoS$_2$ and $10\times$ for GaAs on the grids considered here. These estimates are conservative: the internal momentum grid used to evaluate the independent-particle response is kept fixed, whereas conventional BSE implementations often refine it together with the BSE sampling, further increasing the cost of dielectric screening.

The freedom to choose both the transition-momentum grid and the transferred momentum $\mathbf Q$ also enables calculations tailored to different excitonic properties. For monolayer hBN and MoS$_2$, we resolve the long-wavelength exciton dispersions at transferred momenta about seven times smaller than those accessible on the corresponding uniform meshes, recovering the nonanalytic splitting of low-energy excitons induced by long-range electron--hole exchange. Each point is obtained by reconstructing and solving a new BSE Hamiltonian.  For bulk GaAs, locally refined transition-momentum sampling converges the binding energy of the threefold-degenerate lowest exciton to approximately $4.0$~meV with only a few thousand BSE points, substantially outperforming uniform meshes. This agrees closely with experiment and previous first-principles calculations.

The ability to reconstruct excitonic states at arbitrary momenta within a common smooth electronic representation also provides a basis for combining them consistently with other interaction vertices. This opens the way to first-principles treatments of exciton--phonon coupling~\cite{Alvertis_2024,Marini_Calandra_Cudazzo_2024}, excitonic corrections to nonlocal electron--phonon vertices~\cite{Guandalini_2025}, and nonequilibrium exciton dynamics~\cite{Betancur_2026}.

\begin{acknowledgments}
Funded by the European Union (ERC, DELIGHT,
101052708). Views and opinions expressed are however those of the authors only and do not necessarily
reflect those of the European Union or the European
Research Council. Neither the European Union nor the
granting authority can be held responsible for them.

We are grateful for fruitful discussions with Dr. Iurii Timrov and Dr. Nicola Colonna.
\end{acknowledgments}

\section*{Data Availability Statement}

The data underlying the results presented in this work will be made publicly available upon publication.

\appendix

\section{Static dielectric screening}
\label{subsec:static_dielectric_screening}

The screened interaction entering the direct kernel of Eq.~\eqref{eq:direct_kernel} requires the static microscopic dielectric response. The independent-particle density-density response function can be written in reciprocal space as
\begin{equation}
\label{eq:chi0}
\begin{aligned}
\chi^0_{\mathbf G\mathbf G'}(\mathbf q,\omega)
=
\frac{1}{N_k\Omega}
\sum_{\mathbf k nm} &
\frac{
f_{\mathbf kn}
-
f_{\mathbf k+\mathbf q m}
}{
\omega
+
\varepsilon_{\mathbf kn}
-
\varepsilon_{\mathbf k+\mathbf q m}
+
i\eta
}
\\
&\times
\rho^{\mathbf G'}_{mn}(\mathbf k,\mathbf q)
\left[
\rho^{\mathbf G}_{mn}(\mathbf k,\mathbf q)
\right]^* ,
\end{aligned}
\end{equation}
where $f_{\mathbf kn}$ denotes the occupation of the state $(\mathbf k,n)$ and the transition densities are defined in Eq.~\eqref{eq:rho_periodic}. Equation~\eqref{eq:chi0} involves a sum over the internal $\mathbf k$-point sampling, distinct from the $\mathbf q$ mesh on which the dielectric response is evaluated.

We denote matrices in reciprocal-lattice-vector space by a hat, such that for a generic quantity $A_{\mathbf G\mathbf G'}$,
\begin{equation}
\left[\hat A\right]_{\mathbf G\mathbf G'}
=
A_{\mathbf G\mathbf G'}.
\end{equation}
In particular, the bare Coulomb interaction is represented by the diagonal matrix
\begin{equation}
\left[\hat v(\mathbf q)\right]_{\mathbf G\mathbf G'}
=
v_{\mathbf G}(\mathbf q)\,
\delta_{\mathbf G\mathbf G'}.
\end{equation}
Unless otherwise stated, products and inverses of hatted quantities in the following equations are understood as matrix products and matrix inverses in reciprocal-lattice-vector space. 
Starting from $\hat\chi^0$, the interacting density response satisfies the Dyson equation~\cite{Adler_1962,Wiser_1963,Onida_2002}
\begin{equation}
\hat\chi(\mathbf q,\omega)
=
\hat\chi^0(\mathbf q,\omega)
+
\hat\chi^0(\mathbf q,\omega)
\left[
\hat v(\mathbf q)+\hat f_{\mathrm{xc}}(\mathbf q,\omega)
\right]
\hat\chi(\mathbf q,\omega),
\end{equation}
where $\hat f_{\mathrm{xc}}$ is the exchange--correlation kernel. Within the random-phase approximation, adopted throughout this work, $\hat f_{\mathrm{xc}}=0$, yielding
\begin{equation}
\hat\chi^{\mathrm{RPA}}(\mathbf q,\omega)
=
\left[
\hat{\mathbf 1}
-
\hat\chi^0(\mathbf q,\omega)\hat v(\mathbf q)
\right]^{-1}
\hat\chi^0(\mathbf q,\omega).
\end{equation}
The symmetrised inverse microscopic dielectric matrix is then
\begin{equation}
\hat\epsilon^{-1}(\mathbf q,\omega)
=
\hat{\mathbf 1}
+
\hat v^{1/2}(\mathbf q)
\hat\chi(\mathbf q,\omega)
\hat v^{1/2}(\mathbf q),
\end{equation}
where $\hat v^{1/2}$ is the diagonal matrix with elements $\sqrt{v_{\mathbf G}(\mathbf q)}$. The screened Coulomb interaction follows as
\begin{equation}
\label{eq:W_from_epsilon}
\hat W(\mathbf q,\omega)
=
\hat v^{1/2}(\mathbf q)
\hat\epsilon^{-1}(\mathbf q,\omega)
\hat v^{1/2}(\mathbf q),
\end{equation}
or, equivalently,
\begin{equation}
\hat W(\mathbf q,\omega)
=
\hat v(\mathbf q)
+
\hat v(\mathbf q)
\hat\chi(\mathbf q,\omega)
\hat v(\mathbf q).
\end{equation}
The matrix elements of $\hat W$ are the quantities $W_{\mathbf G\mathbf G'}$ entering Eq.~\eqref{eq:direct_kernel}. As customary in the static BSE approximation~\cite{Rohlfing_2000,Onida_2002}, the screening is evaluated at zero frequency,
\begin{equation}
\hat W(\mathbf q)\equiv\hat W(\mathbf q,\omega=0).
\end{equation}

\section{Bare Coulomb interaction}
\label{sec:bare_coulomb}

The bare Coulomb interaction is evaluated analytically at the required momentum. For three-dimensional periodic systems, its reciprocal-space form is
\begin{equation}
v^{3\mathrm D}_{\mathbf G}(\mathbf q)
=
\frac{8\pi}
{|\mathbf q+\mathbf G|^2}.
\end{equation}
For two-dimensional systems in a repeated-slab geometry, we instead employ a Coulomb interaction truncated along the non-periodic direction~\cite{Rozzi_2006,Ismail-Beigi_2006}. Defining
\begin{equation}
\mathbf K=\mathbf q+\mathbf G
\end{equation}
and its components perpendicular and parallel to the layer as
\begin{equation}
K_{\perp}
=
\mathbf K\cdot\hat{\mathbf n},
\qquad
K_{\parallel}
=
\sqrt{|\mathbf K|^2-K_{\perp}^2},
\end{equation}
where $\hat{\mathbf n}$ is the slab normal, the truncated interaction is
\begin{equation}
v^{\mathrm{slab}}_{\mathbf G}(\mathbf q)
=
\frac{8\pi}{|\mathbf K|^2}
\left[
1-
e^{-K_{\parallel}L}
\cos(K_{\perp}L)
\right].
\end{equation}
Here, $L$ is the truncation length, equal to half the supercell extent along the non-periodic direction for the symmetric slab cutoff used here.

\section{Reciprocal-covariant RBF--TPS interpolation}
\label{subsec:interpolation_scheme}

Following the direct Wannier interpolation approach in the optimally smooth subspace (OSS) of Ref.~\cite{Volpato_2025}, the reciprocal-space interpolation scheme can be chosen independently of the Wannier construction. Here, we interpolate the smooth OSS wave-function components $u^{\mathrm{OSS}}_{\mathbf kn}(\mathbf G,\sigma)$ and screening response $f_{\mathbf G\mathbf G'}(\mathbf q)$ using local radial basis functions (RBFs) with the thin-plate-spline (TPS) kernel of Refs.~\cite{Duchon_1977,Micchelli_1986}. This avoids an intermediate transformation to real space and does not require a rectangular tensor-product grid, making the method well suited to non-orthogonal reciprocal lattices and arbitrary sampling within the first Brillouin zone.

The reciprocal-space quantities considered here are not scalar periodic functions of crystal momentum. A reciprocal-lattice translation $\mathbf G''$ of the momentum is accompanied by a corresponding shift of the Fourier indices,
\begin{align}
u^{\mathrm{OSS}}_{\mathbf k+\mathbf G'' n}(\mathbf G,\sigma)
&=
u^{\mathrm{OSS}}_{\mathbf kn}(\mathbf G+\mathbf G'',\sigma),
\label{eq:rbf_wfc_covariance}\\
f_{\mathbf G\mathbf G'}(\mathbf q+\mathbf G'')
&=
f_{\mathbf G+\mathbf G'',\mathbf G'+\mathbf G''}(\mathbf q).
\label{eq:rbf_screening_covariance}
\end{align}
These covariance relations are used to augment the source sampling with neighbouring reciprocal-lattice images. We consider translations
\begin{equation}
\mathbf G''
=
\sum_{\mu\in\mathcal A}
m_\mu\mathbf b_\mu,
\qquad
m_\mu\in\{-1,0,1\},
\end{equation}
where $\mathbf b_\mu$ are reciprocal-lattice vectors and $\mathcal A$ contains the sampled reciprocal directions.

Only images compatible with the retained plane-wave basis are included. For the wave functions, the shifted component $\mathbf G+\mathbf G''$ must be available, while for the screening response both $\mathbf G+\mathbf G''$ and $\mathbf G'+\mathbf G''$ must be retained. The interpolation neighbourhood may therefore depend on the Fourier component being considered.

\begin{figure}[!t]
\centering
\includegraphics[width=0.99\linewidth]{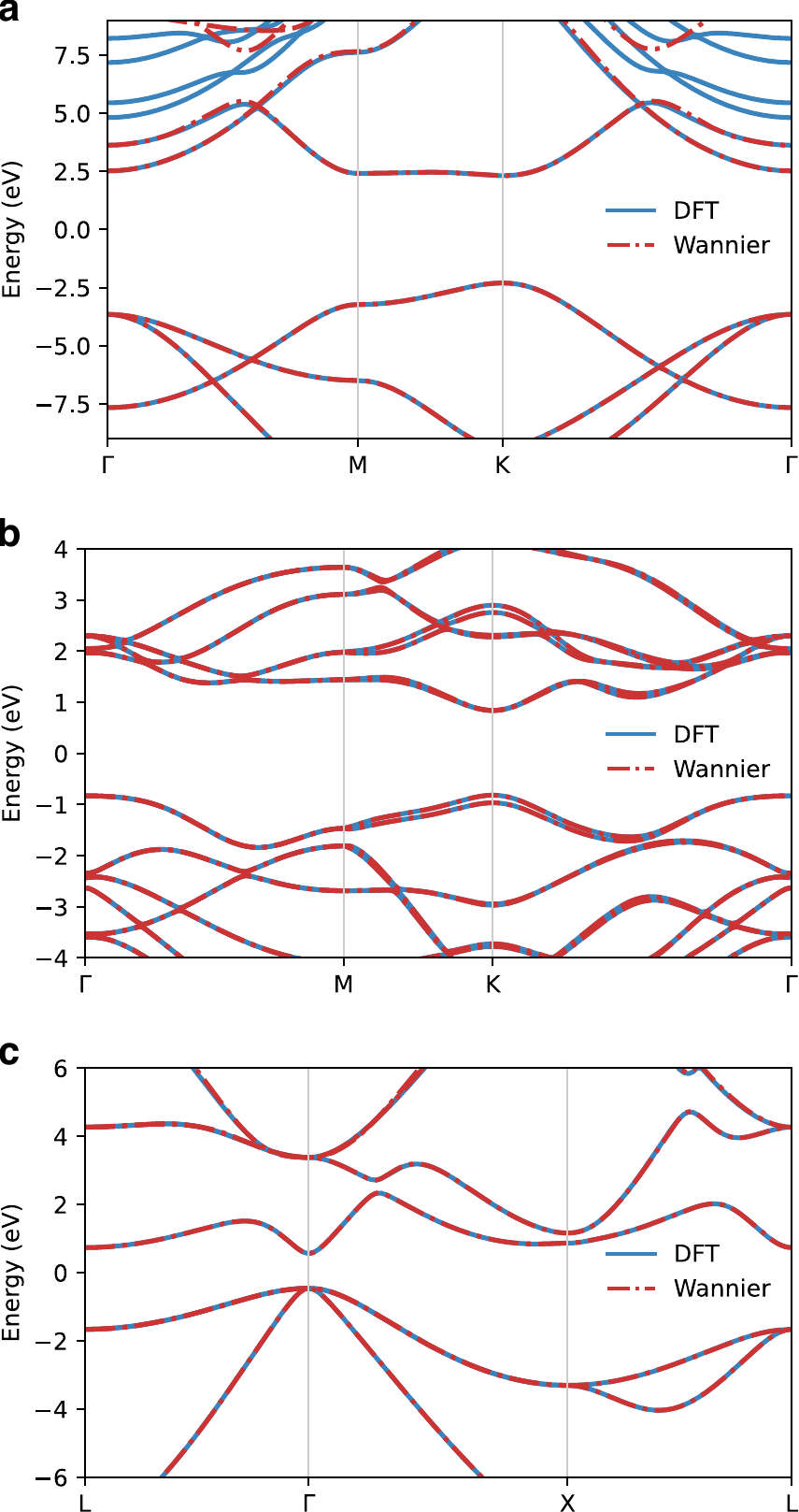}
\caption{\label{fig:wannier_bands}
Comparison between the DFT band structures (blue solid lines) and their Wannier-interpolated counterparts (red dashed lines) for (a) monolayer hBN, (b) monolayer MoS$_2$, and (c) bulk GaAs. The bands shown include the valence and conduction states relevant to the BSE calculations. The Wannier interpolation closely reproduces the DFT dispersions along the $\Gamma$--$M$--$K$--$\Gamma$ path for hBN and MoS$_2$ and the $L$--$\Gamma$--$X$--$L$ path for GaAs. In each case, the centre of the band gap is set to zero energy.}
\end{figure}

For a target momentum $\mathbf x$ in the first Brillouin zone, representing either a Bloch momentum $\mathbf k$ or a screening momentum $\mathbf q$, we first fix the Fourier component to be interpolated: $\mathbf G$ for the wave functions or $(\mathbf G,\mathbf G')$ for the screening response. The source momenta $\mathbf x_i$ are supplemented by their admissible images $\mathbf x_i+\mathbf G''$, from which a local set is selected according to the Cartesian reciprocal-space distance from the target. Labelling each source image by $a=(i,\mathbf G'')$, we define
\begin{equation}
\Delta\mathbf x_a
=
\mathbf x_i+\mathbf G''-\mathbf x
\end{equation}
and the scaled local coordinates
\begin{equation}
\mathbf y_a
=
\frac{\Delta\mathbf x_a}{s},
\qquad
s=
\max_{a\mu}|\Delta x_{a\mu}|.
\end{equation}
We use the RBF kernel
\begin{equation}
\phi(r)
=
\begin{cases}
r^2\ln r, & r>0,\\
0, & r=0,
\end{cases}
\end{equation}
together with a quadratic polynomial augmentation,
\begin{equation}
\mathbf p(\mathbf y)
=
\left(
1,
y_1,\ldots,y_d,
\{y_\mu y_\nu\}_{1\leq\mu\leq\nu\leq d}
\right)^{\mathsf T},
\end{equation}
where $d=2$ or $3$ is the dimensionality of the reciprocal-space sampling. For a local set of $N_{\mathrm{loc}}$ source images, the RBF matrix is
\begin{equation}
\Phi_{ab}
=
\phi\!\left(
\lVert\mathbf y_a-\mathbf y_b\rVert
\right).
\end{equation}
The interpolation weights $\mathbf h$ and the auxiliary coefficients $\boldsymbol\lambda$ associated with the polynomial constraints are obtained from
\begin{equation}
\begin{pmatrix}
\Phi & P\\
P^{\mathsf T} & 0
\end{pmatrix}
\begin{pmatrix}
\mathbf h\\
\boldsymbol\lambda
\end{pmatrix}
=
\begin{pmatrix}
\boldsymbol\phi_*\\
\mathbf p(\mathbf 0)
\end{pmatrix},
\label{eq:reciprocal_rbf_projection}
\end{equation}
where
\begin{equation}
P_{a\ell}=p_\ell(\mathbf y_a),
\qquad
(\boldsymbol\phi_*)_a
=
\phi(\lVert\mathbf y_a\rVert).
\end{equation}
The coefficient $h_a$ is the interpolation weight associated with source image $a$, whereas $\boldsymbol\lambda$ enforces the polynomial-reproduction constraints and does not enter directly in the interpolation of the physical quantity. The quadratic augmentation ensures exact reproduction of polynomials up to second order.

Once the geometric weights are known, they are applied to the reciprocal-space quantities using the covariance relations of Eqs.~\eqref{eq:rbf_wfc_covariance} and~\eqref{eq:rbf_screening_covariance}. For a fixed wave-function component,
\begin{equation}
u^{\mathrm{OSS}}_{\mathbf x n}(\mathbf G,\sigma)
=
\sum_{(i,\mathbf G'')\in\mathcal S_{\mathbf G}(\mathbf x)}
h_{i\mathbf G''}(\mathbf x)\,
u^{\mathrm{OSS}}_{\mathbf k_i n}
(\mathbf G+\mathbf G'',\sigma),
\end{equation}
while for a fixed screening component,
\begin{equation}
f_{\mathbf G\mathbf G'}(\mathbf x)
=
\sum_{(i,\mathbf G'')\in\mathcal S_{\mathbf G\mathbf G'}(\mathbf x)}
h_{i\mathbf G''}(\mathbf x)\,
f_{\mathbf G+\mathbf G'',\,\mathbf G'+\mathbf G''}(\mathbf q_i).
\end{equation}
Here, $\mathbf k_i$ and $\mathbf q_i$ denote the source momenta for the wave-function and screening interpolations, respectively, while $\mathbf G''$ identifies the reciprocal-lattice translation of each source image. The sets $\mathcal S_{\mathbf G}$ and $\mathcal S_{\mathbf G\mathbf G'}$ contain only images for which the required shifted Fourier components are available. Since the weights depend only on the geometry of the corresponding local set, the same interpolation is applied independently to the real and imaginary parts of complex quantities.

\section{Wannier-interpolated band structures}
\label{app:wannier_quality}

Figure~\ref{fig:wannier_bands} compares the DFT and Wannier-interpolated band structures for the three materials considered in this work, including the valence and conduction manifolds relevant to the BSE calculations. For hBN, the initial projections consist of B and N $sp^2$ and $p_z$ states together with two interstitial $s$-like functions. For MoS$_2$, Mo $d$ and S $p$ states are used, while for GaAs the projections include Ga and As $s$ and $p$ states together with $d_{xy}$, $d_{xz}$, and $d_{yz}$ states on both atomic species.

In all cases, the Wannier-interpolated bands closely reproduce the DFT dispersions along the selected high-symmetry paths, including the band-edge regions relevant to the electron--hole transition space. This confirms the accuracy of the Wannier representation used to reconstruct the electronic states entering the BSE at arbitrary $\mathbf k$ points.

\bibliography{bibliography}

\end{document}